\documentclass[]{tJOT2e}
\usepackage[dvips]{color}
\begin{document}
	\title{Geometrical statistics of the vorticity vector and the strain rate tensor in rotating turbulence}
	\author{Lorenzo Del Castello$^{\rm ab}$$^{\ast}$\thanks{$^\ast$Corresponding author. Email: lorenzodelcastello@gmail.com\vspace{6pt}}
			and Herman J.H. Clercx$^{\rm a}$\\
			\vspace{6pt}$^{\rm a}${\em{Department of Physics and J.M. Burgers Centre for Fluid Dynamics, Eindhoven University of Technology, P.O. Box 513, 5600 MB Eindhoven, The Netherlands;}}\\
						$^{\rm b}${\em{Istituto Sistemi Complessi, Consiglio Nazionale delle Ricerche,\\UOS Sapienza, 00185 Rome, Italy}}}
	\maketitle
	\begin{abstract}
		We report results on the geometrical statistics of the vorticity vector obtained from experiments in electromagnetically forced rotating turbulence. A range of rotation rates $\Omega$ is considered, from non-rotating to rapidly rotating turbulence with a maximum background rotation rate of $\Omega=5$ rad/s (with Rossby number much smaller than unity). Typically, the Taylor-scale Reynolds number in our experiments is around ${\rm{Re}}_{\lambda}\approx 100$. The measurement volume is located in the centre of the fluid container above the bottom boundary layer, where the turbulent flow can be considered locally statistically isotropic and horizontally homogeneous for the non-rotating case, see van Bokhoven {\it{et al.}}, {\it{Phys. Fluids}} {\bf{21}}, 096601 (2009).
		Based on the full set of velocity derivatives, measured in a Lagrangian way by 3D Particle Tracking Velocimetry, we have been able to quantify statistically the effect of system rotation on several flow properties. For the range of rotation rates considered, the experimental results show how the turbulence evolves from almost isotropic 3D turbulence ($\Omega\lesssim 0.2$ rad/s) to quasi-2D turbulence ($\Omega\approx 5.0$ rad/s) and how this is reflected by several statistical quantities. In particular, we have studied the orientation of the vorticity vector with respect to the three eigenvectors of the local strain rate tensor and with respect to the vortex stretching vector.
		Additionally, we have quantified the role of system rotation on the self-amplification terms of the enstrophy and strain rate equations and the direct contribution of the background rotation on these evolution equations. The main effect of background rotation is the strong reduction of extreme events and related (strong) reduction of the skewness of PDFs of several quantities such as, for example, the intermediate eigenvalue of the strain rate tensor and the enstrophy self-amplification term.
	\end{abstract}
	\section{Introduction}\label{Sec1}
		The effect of the background rotation on the dynamics of fluid flows is ubiquitous in large-scale geophysical and astrophysical flows, as well as in the context of industrial rotating machinery. It is well-known that the Coriolis acceleration term in the Navier--Stokes equations is responsible for altering the flow dynamics including the anisotropisation of turbulent flows, but the underlying physical mechanisms are still poorly understood. Another familiar phenomenon is that a three-dimensional (3D) turbulent flow subject to a fast background rotation evolves towards a quasi-two-dimensional (Q2D) state, which is characterised by a strong damping of velocity gradient components along the direction parallel to the rotation axis.\\
		The anisotropisation of turbulent flows induced by background rotation, has been the subject of several numerical and experimental investigations in the past, which led to important progress in the field. We briefly summarise the most important studies based on simulations and physical experiments, and the main observations. Early laboratory experiments of rotating grid-turbulence were carried out in a wind tunnel~\cite{traugott1958naca,wigeland1978iit,jacquin1990jfm} and focused, for example, on the decay of the kinetic energy and the energy dissipation rate~\cite{traugott1958naca} and on the confirmation of the role of the nonlinear nature of the transition from 3D to predominantly 2D flow dynamics of homogeneous turbulence (predicted earlier by Cambon {\it{et al.}}~\cite{cambon1989jfm}).
		The increase of Eulerian velocity correlations due to rotation and the temporal decay of the turbulent flow were already quantified during the seventies with an experiment where turbulent air flow in a rotating annular container was forced by a system of translating grids~\cite{ibbetson1975jfm}. In 1976 McEwan~\cite{mcewan76} revealed for the first time the concentration of vorticity in coherent structures in rotating turbulence. Hopfinger and colleagues~\cite{hopfinger1982jfm} investigated the large-scale effects of rotation on a turbulent flow, which was continuously forced locally in space, aimed at studying the population statistics of the vorticity tubes characterising the rotating flow. They also provided a detailed phenomenological description of the instabilities of such eddies for a specific rotation rate, their nonlinear mutual interactions and eventual breakdowns.
		During the last two decades, Direct Numerical Simulation (DNS) of rotating turbulence (decaying, or with large-scale forcing) revealed additional information.
		For example, the important increase of velocity correlations along the direction parallel to the rotation axis, and the mild decrease of correlations along the perpendicular directions, with increasing system rotation~\cite{yeung1998pof}, and the increase of horizontal integral length scales with increasing rotation rate, followed by a decrease of the same horizontal integral scales for the fastest rotation rates~\cite{godeferd1999jfm}. The latter DNS focused on the combined effects of the background rotation and the vertical (top and bottom) confinement on turbulence forced locally in space.
		Recent laboratory experiments on rotating turbulence addressed several issues like intermittency~\cite{baroud2003pof}, coupling between inertial wave patterns, and turbulence decay using high-resolution Particle Image Velocimetry (PIV)~\cite{morize2005pof,morize2006pof,moisy2010jfm}, and accurate visualisations by means of reflective flakes of the formation and evolution of columnar eddies in rotating turbulence~\cite{davidson2006jfm}.
		These visualisation experiments showed that -- for initially inhomogeneous turbulence -- large coherent vortices build-up in a time comparable with half the revolution period, compatible with linear effects, rather than on the longer time scale typical of nonlinear ones. The stereo-PIV measurements by van Bokhoven and coworkers~\cite{bokhoven2009pof} characterised the effects of rapid background rotation on the statistical properties of the same turbulent flow reported in the present paper. They explored the statistical homogeneity and (planar) isotropy of the velocity fluctuations, and extracted the integral time and length scales as function of rotation rate. In particular, they described, for the first time in laboratory settings, the reverse dependence on the rotation rate of the spatial horizontal correlation coefficients, as predicted earlier by numerical studies~\cite{godeferd1999jfm}.
		Furthermore, they observed a linear (anomalous) scaling of the longitudinal spatial structure function exponents in the presence of rotation. With the present-day computer resources, a surge in large-scale DNS of rotating turbulence has occurred, although mostly for periodic domains (thus excluding the effects of confinement and in absence of Ekman boundary-layer dynamics close to the domain walls)~\cite{thalabard2011,mininni2012}.\\
		The experimental data available is still scarce and mostly of Eulerian nature. Recently, few studies have been reported on the three velocity components of the flow measured in a Lagrangian manner by using 3D Particle Tracking Velocimetry (3D-PTV)~\cite{kinzel2010,delcastello2011pre,delcastello2011prl}. Two of these studies were carried out in our laboratory, and focused on the Lagrangian statistics of passive tracer velocities and accelerations in rotating turbulence for a sequence of rotation rates~\cite{delcastello2011pre,delcastello2011prl}. Thanks to 3D-PTV, we have access to the vorticity vector, and studied its dynamics in rotating turbulence. For the non-rotating case, this kind of particle tracking studies has been pursued several years ago~\cite{luethi2005jfm}, and rotating PTV experiments quantified the transition from 3D to Q2D large-scale dynamics~\cite{luethi2008ercoftac}.
		But laboratory experiments aimed at exploring the influence of system rotation on  the dynamics of the small scales and in particular of the vorticity vector have not yet been reported and are the main topics of the present paper.
		In the context of the existing literature, the latter studies on the Lagrangian statistics of velocities and accelerations of passive tracers, and the present work reported here on the geometrical statistics of the vorticity and the strain rate tensor in rotating turbulence are based on experiments resembling the ones performed in closed non-shallow containers, and with continuous forcing applied locally in space (see, e.g., Refs.~\cite{hopfinger1982jfm,davidson2006jfm}). The forcing scheme adopted to continuously sustain the turbulence produces a flow which is similar to a Taylor-Green flow, resembling the driving mechanism used in many DNS simulations of turbulence (see, for example, Ref.~\cite{mininni2009pof}).\\
		The present paper is organised as follows. In Section~\ref{Sec2}, the theoretical background of geometrical statistics of the vorticity vector is briefly summarised. The experimental set-up is described briefly in Section~\ref{Sec3}, while more detailed information can be found elsewhere~\cite{delcastello2010phd,delcastello2011pre,delcastello2011prl}. Based on our stereo-PIV~\cite{bokhoven2009pof} and 3D-PTV experiments, the main features of forced rotating turbulence relevant for the present study is discussed in Section~\ref{Sec4}. Results from the study on some geometrical statistics of the flow will be presented in Sections~{\ref{Sec5}} (the reference non-rotating case) and in Section~\ref{Sec6} for the cases when background rotation is applied. Finally, conclusions are summarised in Section~{\ref{Sec7}}.
	\section{Theoretical background on the geometrical statistics of vorticity}\label{Sec2}
		The governing equations to describe incompressible flows in a fluid subject to system rotation are usually the Navier-Stokes equations formulated for the rotating non-inertial frame of reference. They contain two additional contributions, the Coriolis acceleration and the centrifugal acceleration. The latter is irrotational, therefore can be written as a gradient and is incorporated in the pressure gradient term. For a Newtonian fluid, and in the presence of external forces, the Navier-Stokes equations read (in tensorial notation):
		\begin{equation}\label{eq:NSmass-rotating}
			\frac{\partial u_i}{\partial x_i} = 0~,
		\end{equation}
		\begin{equation}\label{eq:NSmomentum-rotating}
			\frac{Du_i}{Dt} \equiv \frac{\partial u_i}{\partial t} + u_j\frac{\partial u_i}{\partial x_j} = 2\epsilon_{ijk}u_j\Omega_k -\frac{1}{\rho}~\frac{\partial p}{\partial x_i} + \nu\frac{\partial^2u_i}{\partial x_j\partial x_j} + F_i~.
		\end{equation}
		Here, $u_i$ are the components of the velocity ${\bf{u}}$ (for later use we also introduce the notation ${\bf{u}}=(u,v,w)$), $x_i$ the components of the position vector ${\bf{x}}=(x,y,z)$, and ${\partial u_i}/{\partial x_j}$ the components of the velocity gradient tensor ${\bf{\nabla}}{\bf{u}}$. The components of the system rotation ${\bf{\Omega}}$ are denoted by $\Omega_i$ and those of the external force field ${\bf{F}}$ applied to the fluid by $F_i$. Keeping the laboratory experiments in mind, the system rotation is defined such that only the vertical component is non-zero, i.e. $\Omega_3 = |{\bf{\Omega}}| = \Omega$. Finally, $\nu$ is the kinematic viscosity of the fluid, and $\rho$ its density.\\
		It is also convenient to define here the nondimensional numbers which characterise a fluid flow according to the relative importance of the different terms in the momentum equation. Indicating with $\mathcal{L}$ and $\mathcal{U}$ the typical length and velocity scales representative of the flow (such as the integral length scale and the root mean square (rms) velocity of the turbulent flow, respectively), the ratio between the order of magnitude of different terms in Eq. (\ref{eq:NSmomentum-rotating}) defines three relevant dimensionless parameters. The first is the Reynolds number ${\rm{Re}} = {\mathcal{U}}{\mathcal{L}}/\nu$. The Rossby number ${\rm{Ro}} = {\mathcal{U}}/({2\Omega\mathcal{L}})$ measures the relative importance of the Coriolis force with respect to fluid inertia. For non-rotating flows ${\rm{Ro}}=\infty$, and for rapidly rotating flows we have ${\rm{Ro}}\rightarrow 0$.
		Finally, the Ekman number ${\rm{Ek}} = {\nu}/({\Omega \mathcal{L}^2})$, which is a measure for the relative importance of viscous forces with respect to the Coriolis force.\\
		For ${\rm{Ek}}\ll 1$ (in the bulk of the flow, away from boundaries) and ${\rm{Ro}}\ll 1$, the viscous and advective terms may be neglected, and in steady conditions the fluid particle acceleration is solely determined by the pressure gradient and the Coriolis force. Such situation is known as the geostrophic balance, and it is of utmost importance for the dynamics of the flow in the atmosphere and in the oceans. From such an expression, the Taylor-Proudman theorem can easily be derived: $\left({\bf{\Omega}}\cdot{\bf{\nabla}}\right){\bf{u}}={\bf{0}}$. This theorem states the suppression of the velocity derivatives in the direction parallel with the rotation axis.\\
		In order to understand the interactions between the terms of the equations of motion ruling a turbulent flow, it is useful to express them as functions of the vorticity vector $\omega_i=\epsilon_{ijk}\partial u_k/\partial x_j$. The vorticity evolution equation in the rotating frame of reference, derived by taking the curl of the Navier-Stokes equations (\ref{eq:NSmomentum-rotating}), reads:
		\begin{equation}
			\frac{D\omega_{a,i}}{Dt}=\omega_{a,j}\frac{\partial u_i}{\partial x_j}+
				\nu\nabla^2\omega_{a,i}+\varepsilon_{ijk}\frac{\partial F_k}{\partial x_j}~,
			\label{absvor-eq}
		\end{equation}
		with the absolute vorticity defined according to $\omega_{a,i} = \omega_i + 2\Omega_i$ (note that ${\bf{\Omega}}$ is independent of both space and time).\\
		Before we discuss in more detail the role of system rotation on vorticity dynamics, it is instructive to summarise the main results for homogeneous isotropic turbulence. It had already been recognised for more than 70 years, see for example Ref.~\cite{taylor1938}, that the velocity derivatives in the turbulent flow play a special role in vorticity dynamics in general, and in vortex stretching and tilting processes in particular. In the case of absence of system rotation, the vorticity equation (\ref{absvor-eq}) reduces to 
		\begin{equation}
			\frac{D\omega_i}{Dt}=\omega_j s_{ij}+\nu\nabla^2\omega_i+\varepsilon_{ijk}\frac{\partial F_k}{\partial x_j}~,
		\end{equation}
		where $s_{ij}=\frac{1}{2}\left(\partial u_i/\partial x_j + \partial u_j/\partial x_i\right)$ is the strain rate tensor (the symmetric part of the velocity gradient tensor). The term $\omega_js_{ij}=\omega_j\partial u_i/\partial x_j$ represents the stretching (or compression) and tilting of the vorticity vector operated by the velocity gradient. Since the work by Taylor~\cite{taylor1938}, the vortex stretching operated by the strain rate field is considered as a key--ingredient in the dissipative process of turbulence. The evolution of the local enstrophy, $V=\frac{1}{2}\omega^2 = \frac{1}{2}\omega_i\omega_i$, and the local strain, $s^2=s_{ij}s_{ij}$, are expressed by the following equations:
		\begin{equation}
			\frac{DV}{Dt} = \frac{1}{2}\frac{D\omega^2}{Dt} = \omega_i\omega_j s_{ij} + \nu \omega_i \nabla^2\omega_i + 
				\varepsilon_{ijk}\omega_i\frac{\partial F_k}{\partial x_j}~, 
			\label{vort-prod}
		\end{equation}
		and
		\begin{equation}
			\frac{1}{2}\frac{Ds^2}{Dt} = -s_{ij}s_{jk}s_{ki} - \frac{1}{4}\omega_i\omega_j s_{ij} - 
				s_{ij}\frac{1}{\rho}\frac{\partial^2 p}{\partial x_i\partial x_j} + \nu s_{ij}\nabla^2 s_{ij} + s_{ij}F_{ij}~,
			\label{strain-prod}
		\end{equation}
		with $F_{ij} =\frac{1}{2}\left(\partial F_i/\partial x_j + \partial F_j/\partial x_i\right)$. The evolution of the vorticity and strain fields are strongly connected, see Refs.~\cite{tsinober2001,luethi2005jfm}. In particular, the dynamics of both fields are driven by self-amplification. As discussed by Tsinober and coworkers~\cite{galantitsinober2000,tsinober2001}, the production terms of enstrophy and strain are dominated by the terms $\omega_i\omega_js_{ij}$ and $s_{ij}s_{jk}s_{ki}$, respectively. The averaged production terms, in the following denoted by $\langle\omega_i\omega_js_{ij}\rangle$ and $\langle -s_{ij}s_{jk}s_{ki}\rangle$, respectively, have also been observed to be larger than the external forcing term by several orders of magnitude in numerically simulated turbulent flows. See, for example, Ref.~\cite{galantitsinober2000} and Table II in that paper.
		The numerical data revealed that both $\langle\omega_i\omega_js_{ij}\rangle\ge 0$ and $\langle -s_{ij}s_{jk}s_{ki}\rangle\ge 0$.
		In fact, the self-amplification term was found to be ${\mathcal{O}}(10^2)$ compared to the forcing term in Eq.~(\ref{vort-prod}) at ${\rm{Re}}_{\lambda}=35$; ${\mathcal{O}}(10^3)$ at ${\rm{Re}}_{\lambda}=110$; ${\mathcal{O}}(10^4)$ at ${\rm{Re}}_{\lambda}=250$. These observations were made for the mean (volume-averaged) values, and also point-wise throughout the flow field. As remarked by L{\"{u}}thi {\it {et al.}}~\cite{luethi2005jfm}, indications exist that the self-amplification process may have universal character for a wide range of turbulent flows.\\
		It is important to decompose the interaction between vorticity and strain rate in the eigenframe of $s_{ij}$ to distinguish between stretching, compression, and tilting events. Such analysis has been extensively conducted on experimental turbulence data (\textit{e.g.}, see \cite{saffmann1991,tsinober1992,tsinober1996,kholmyansky2001pof,guala2005jfm,luethi2005jfm}), studying the alignment of the vorticity vector ${\boldsymbol{\omega}}$ with respect to the three eigenvectors ${\boldsymbol{\lambda}}_{\alpha}$ of the strain rate tensor.\\
		In this paper we present the results of the same analysis on experimental data of rotating, confined, continuously forced turbulence, in order to statistically quantify the influence of the background rotation in terms of such mutual alignments.
	\section{The set-up for the rotating turbulence experiment}\label{Sec3}
		\begin{figure}[!h]
			\centering
			\includegraphics[width=0.57\textwidth]{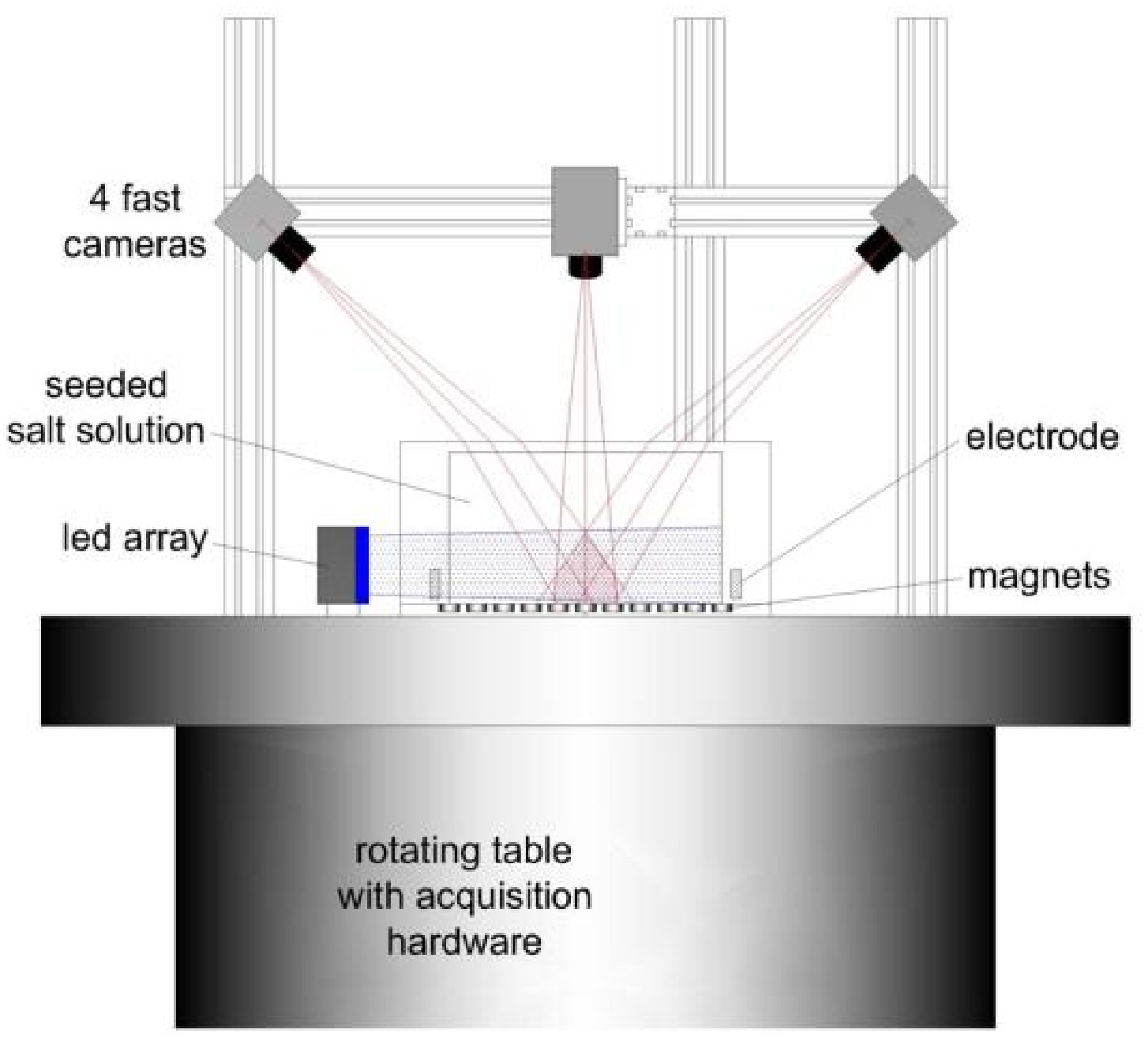}
			\includegraphics[width=0.42\textwidth]{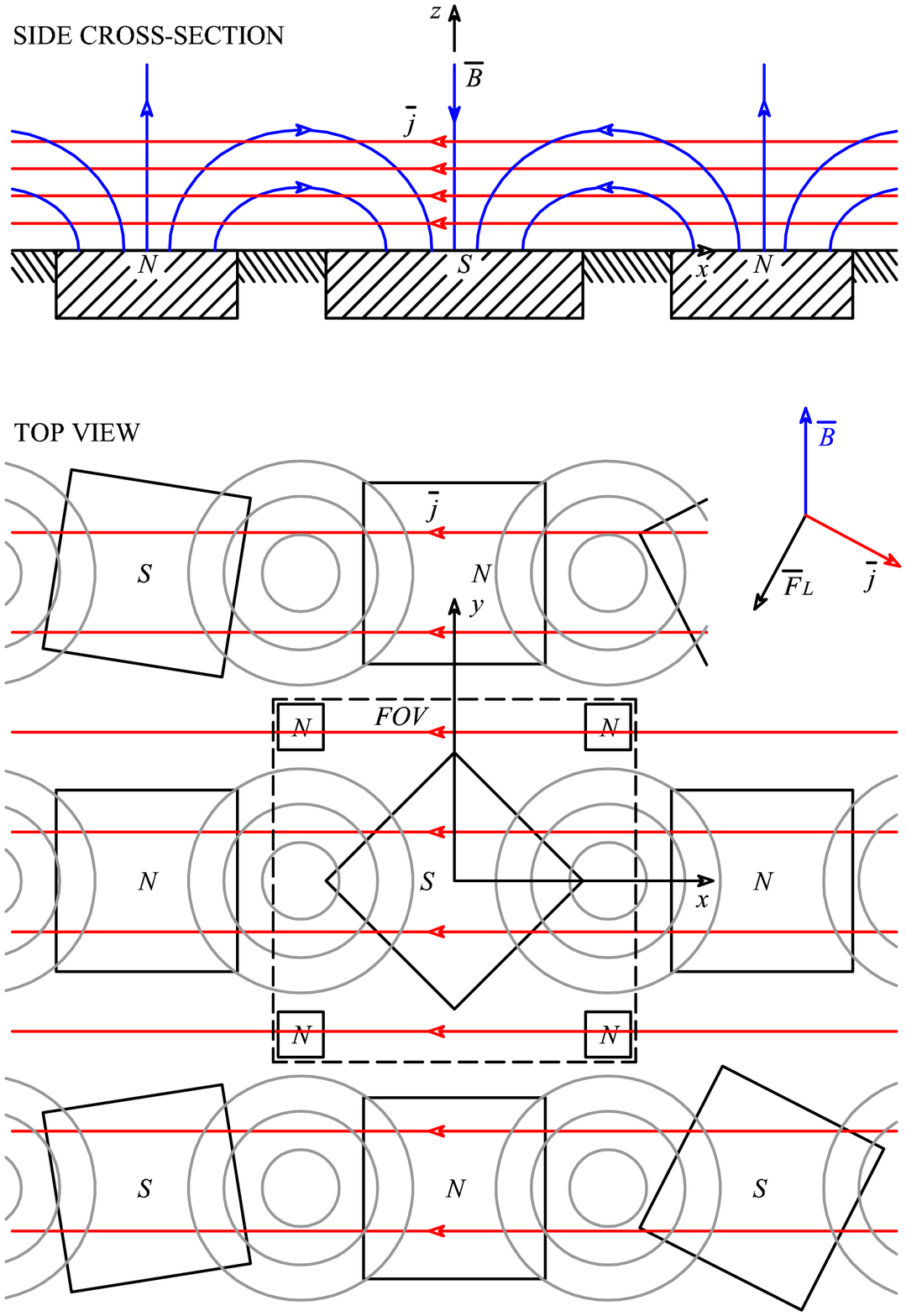}
			\caption{(Colour online) Panel (a): Schematic drawing of the experimental setup, side view. A perspex container sits on top of a rotating table, and is filled with a NaCl solution. An array of permanent magnets is placed below the container, and two linear electrodes are immersed in the fluid. An aluminium frame holds four digital cameras in stable position (three of them are visible in the drawing), and their common field--of--view is sketched. A powerful LED--array, on the left of the container, provides the necessary illumination. Panel (b): Schematic drawing of the forcing system, $xz$-section through the origin of the central part of the forced region of the flow, and top view of the same region. The magnetic field ${\bf{B}}$ and the current density ${\bf{j}}$ are indicated, together with the position of the magnets and their top-face polarity. The measurement region is marked with a dashed line.
			The position of the Cartesian reference frame $\{x,y,z\}$ is indicated in the drawing.}
			\label{fig1}
		\end{figure}
		A transparent fluid container equipped with a turbulence generator and an optical measurement system is used to perform the experiments. A side-view of the setup is sketched in Fig.~\ref{fig1}(a). Four high-speed cameras (Photron FastcamX--1024PCI) picture the central-bottom region of the fluid container through the top-lid. The illumination is provided by a LED--array made of $238$ Luxeon K2 LEDs ($1.4~\mathrm{kW}$ total dissipation and roughly $150~\mathrm{W}$ of light) mounted on a water-cooled aluminium block. The entire setup is mounted on a rotating table, so that the flow is measured in the rotating frame of reference. The flow domain measures $500\times500\times250~\mathrm{mm^3}$ (length $\times$ width $\times$ height), and the free surface deformation is inhibited by a perfectly sealed top-lid.
		The turbulence generator is inspired by the electromagnetic forcing system commonly used for shallow-flows~\cite{sommeria1986,tabeling1991,dolzhanskii1992}, and currently operational in our laboratory for both shallow flow and rotating turbulence experiments~\cite{clercx2003,akkermans2008,bokhoven2009pof}. A schematic sketch of the forcing system is shown in Fig.~\ref{fig1}(a). The tank is filled with sodium chloride (NaCl) water solution ($25~\mathrm{g}$ NaCl in $100~\mathrm{g}$ of water). The fluid density $\rho_{f}$ is $1.19~\mathrm{g/cm^3}$, and the kinematic viscosity $\nu$ is $1.319~\mathrm{mm^2/s}$. Two elongated electrodes, positioned near the bottom at opposite sidewalls of the container, are connected to a remote-controlled power supply (KEPCO BOP 50 8P) which drives a stable electric current ($8.39~\mathrm{A}$) through the fluid.
		Permanent axial magnets (neodymium, maximum magnetic strength $1.4~\mathrm{T}$) are placed underneath the bulk fluid with top-face polarities arranged in a check-board scheme. The schematic sketch of the array of magnets is shown in Fig.~\ref{fig1}(b). The density of the magnetic field lines in the fluid bulk is increased by a steel plate $10~\mathrm{mm}$ thick on which the magnets are sitting. A range of flow scales is forced by using two differently sized magnets: $10\times10\times20~\mathrm{mm^3}$ and $40\times40\times20~\mathrm{mm^3}$~\cite{bokhoven2009pof}.
		The spacing between adjacent large magnets, $\mathcal{L}^F=70~\mathrm{mm}$, represents the energy input length scale in the system. Note that the flow is forced in the bottom region of the domain, and the turbulent kinetic energy decays with the distance $z$ from the bottom wall in the non-rotating case. However, the vertical profiles of turbulent kinetic energy and related quantities appear to be rather flat when the rotation rate $\Omega\geq1.0~\mathrm{rad/s}$~\cite{delcastello2010phd}.\\
		The Particle Tracking Velocimetry (PTV) code developed at ETH, Z\"{u}rich~\cite{maas1993eif1,malik1993eif2,willneff2002istpdrm,willneff2003phd,luethi2005jfm} is used to perform the measurements. PMMA (poly methyl methacrylate) particles (diameter $d_p=127\pm3~\mathrm{\mu m}$; density $\rho_p=1.19~\mathrm{g/cm^3}$) are used as tracers. The concentration of the salt solution is adjusted to match the PMMA density. The Stokes number  $St=\tau_p/\tau_{\eta}$ expresses the ratio between the particle response time and a typical time scale of the flow. Here $\tau_p = (d_p^2 \rho_p)/(18\nu\rho_f)$, $\rho_{p}/\rho_f=1$, and $\tau_{\eta}$ is the Kolmogorov time scale of the turbulent flow ($0.25~{\rm{s}}\lesssim\tau_{\eta}\lesssim0.55~{\rm{s}}$ for different runs of our experiment). Therefore $St={\mathcal{O}}(10^{-3})$, and the chosen seeding particles well approximate passive flow tracers in terms of buoyancy and inertia.
		The system is calibrated using a 3D-target, and the calibration parameters are further optimised using seeded flow images. The 3D--position of the particles is retrieved with a maximum error of $9~\mathrm{\mu m}$ in the horizontal directions, and $18~\mathrm{\mu m}$ in the vertical one. The recovered tracer trajectories are filtered to remove the noise produced by the positioning uncertainty, fitting third--order polynomials along limited segments of the trajectories around each particle position (for details, see Ref.~\cite{luthi2002phd}). From the coefficients of the polynomial in each point, the position, velocity, and acceleration of the tracers at each time--step are extracted. Up to 2500 particles per time--step have been tracked on average in a volume with size $100\times100\times100~\mathrm{mm^3}$, thus roughly $1.5\mathcal{L}^F$ along each coordinate direction.\\
		A detailed description of the experimental setup and the data processing routines, together with an in-depth characterisation of the flow, can be found in Refs.~\cite{bokhoven2009pof,delcastello2010phd}. Experimental settings and some basic flow features relevant for the present experiments are summarised in the following Section.
	\section{The general flow characteristics}\label{Sec4}
		\linespread{1.3} 
		\begin{table}[!b]
			\begin{center}
				\begin{tabular}{lccccccc}
					\hline
					$\Omega$ \footnotesize{(rad/s)} & & 0 & 0.2 & 0.5 & 1.0 & 2.0 & 5.0\\
					\hline
					\hline
					$\gamma_h$ \footnotesize{($\mathrm{mm/s}$)}
						&  & 9.6 & 9.3 &  9.8 & 12.1 & 17.3 & 13.3 \\
					$\gamma_z$ \footnotesize{($\mathrm{mm/s}$)}
						&  & 8.3 & 7.7 &  7.8 &  6.6 &  7.3 &   2.2 \\
					$\xi = \gamma_z/\gamma_h $ (--)	&  & 0.86 &  0.83 &  0.80 &  0.55 &  0.42 &  0.17 \\ 
					\hline
					$Ro = u_{rms}/(2\Omega{\mathcal{L}}^F)$ \footnotesize{(--)}
						& & $\infty$ & 0.47 & 0.20 & 0.13 & 0.09 & 0.02 \\
					$Ek\times 10^5$ \footnotesize{(--)}
						& & $\infty$ & 10 & 4 & 2 & 1 & 0.4 \\
					$\delta_{Ek}$ \footnotesize{($\mathrm{mm}$)}
						& & $\infty$ & 2.5 & 1.6 & 1.1 & 0.8 & 0.5 \\
					\hline\\
				\end{tabular}
				\linespread{1}
				\caption{For each (non) rotating experiment, we have summarised the root-mean-square (rms) values of the horizontal and vertical velocities, $\gamma_h = [(\langle u_x^2\rangle + \langle u_y^2\rangle)/2]^{1/2}$ and $\gamma_z=\langle u_z^2\rangle^{1/2}$, respectively, and the ratio of vertical and horizontal rms values $\xi=\gamma_z/\gamma_h$. The Rossby number $Ro = u_{rms}/(2\Omega{\mathcal{L}}^F)$, the Ekman number $Ek = \nu/(\Omega L_z^2)$, with $L_z = 250~{\rm{mm}}$ the vertical size of the flow domain, and the thickness of the Ekman boundary layer $\delta_{Ek} = \sqrt{\nu/\Omega}$ are also given.}
				\label{tab1}
			\end{center}
		\end{table}\linespread{1}%\\
		For the exploration of the impact of a range of rotation rates on turbulence dynamics, the rotating table is set to spin at different rotation rates $\Omega\in\lbrace0; 0.2; 0.5; 1.0; 2.0; 5.0\rbrace~\mathrm{rad/s}$ around the vertical $z$--axis. The kinetic energy of the flow is statistically steady in time and decays in space along the upward $z$--direction. The turbulent flow is approximately statistically homogeneous in the horizontal directions (and almost statistically isotropic on horizontal planes sufficiently far above the boundary layers), see the stereo-PIV measurements conducted during previous turbulence studies in the same system~\cite{bokhoven2009pof}. The flow is fully turbulent in the bottom region of the container where the PTV measurements are performed. The Eulerian characterisation of the (rotating) turbulent flow with stereo-PIV measurements has been reported in Ref.~\cite{bokhoven2009pof}, which turned out to be extremely useful for validation of our PTV measurements.\\
		In Table~\ref{tab1} we have summarised some of the key quantities of the flow. The transition from a 3D flow to a Q2D one, in first approximation, is quantified in terms of the ratio $\xi = \gamma_z/\gamma_h$ of the vertical ($\gamma_z = \langle u_z^2\rangle^{1/2}$) and horizontal ($\gamma_h = [(\langle u_x^2\rangle + \langle u_y^2\rangle)/2]^{1/2}$) rms velocity fluctuations. Apart from an anomaly in the values of $\gamma_z$ and $\gamma_h$ observed for $\Omega=2.0~\mathrm{rad/s}$, the ratio $\xi$ is seen to decrease monotonically with increasing rotation rate $\Omega$. The vertical velocity at the maximum rotation rate $\Omega=5.0~\mathrm{rad/s}$ is seen to be strongly suppressed ($\xi = \gamma_z/\gamma_h\ll 1$), indicating the presence of a two-dimensionalisation process of the flow field. It is noteworthy to emphasise the higher value of both $\gamma_h$ and $\gamma_z$ for $\Omega=2.0~\mathrm{rad/s}$.
		This anomalous behaviour may be connected with instabilities of large-scale anticyclonic vortical structures at this rotation rate, see Refs.~\cite{kloosterziel1991,vanheijst2009}. This behaviour may be expected for Rossby numbers close to the critical value $0.1$ (${\rm{Ro}}\approx 0.2$ in similar experiments by Hopfinger~\textit{et al.}~\cite{hopfinger1982jfm}). For the Kolmogorov length and time scales we found the typical values $0.6~{\rm{mm}}\lesssim\eta\lesssim0.8~{\rm{mm}}$ and $0.25~{\rm{s}}\lesssim\tau_{\eta}\lesssim0.55~{\rm{s}}$, respectively. The Taylor-scale Reynolds number is found to be in the range $70\lesssim Re_{\lambda}\lesssim110$ for all rotation rates.
	\section{Geometrical statistics in case of no rotation}\label{Sec5}
		\begin{figure}[!b] 
			\centering
			\includegraphics[width = 0.5\textwidth]{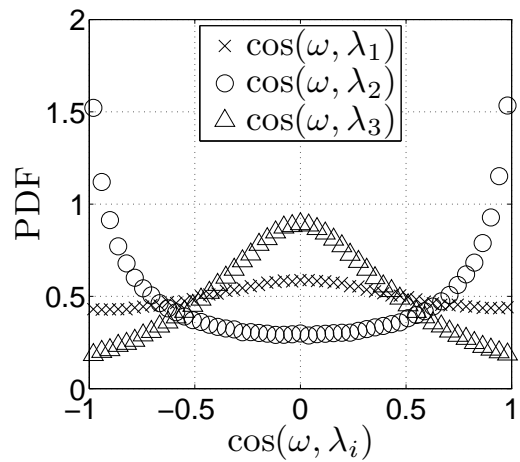}%{exp01bis-0rad_cosomegalambda123-fstep30-linlin}
			\caption{PDFs of the cosine of the angle between the vorticity vector ${\boldsymbol{\omega}}$ and the three eigenvectors ${\boldsymbol{\lambda}}_{\alpha}$ of the strain rate tensor, for the reference non-rotating experiment.}
			\label{fig2}
		\end{figure}
		Already thirty years ago it has been observed, see Refs.~\cite{siggia1981,ashurst1987,tsinober1998a,tsinober2001}, that the vorticity vector in a turbulent flow is predominantly aligned with the intermediate eigenvector of the strain rate tensor. The eigenvectors of the strain rate tensor are denoted by ${\boldsymbol{\lambda}}_{\alpha}$, with $\alpha\in\{1,2,3\}$, and the associated eigenvalues are $\Lambda_{\alpha}$. The eigenvalues are ordered such that $\Lambda_1 > \Lambda_2 > \Lambda_3$, and $\Lambda_1 + \Lambda_2 + \Lambda_3 = 0$ for incompressible flows. It can be shown straightforwardly that the instantaneous strain production term is equivalent to $-s_{ij}s_{jk}s_{ki}=-(\Lambda_1^3+\Lambda_2^3+\Lambda_3^3)$ and, in the case of incompressibility, it is also equal to $-3\Lambda_1\Lambda_2\Lambda_3$~\cite{betchov1956jfm}.
		This immediately implies that the condition $\Lambda_2>0$ is equivalent to strain production (or equivalently, the positiveness of $\langle \Lambda_2\rangle >0$ explains the positiveness of the averaged strain production term $\langle -s_{ij}s_{jk}s_{ki}\rangle$).
		More recently it has been reported by L{\"{u}}thi and coworkers~\cite{luethi2005jfm} that the preferential alignment of the vorticity vector with the eigenvector ${\boldsymbol{\lambda}}_2$ persists even in regions of the flow field where strain and enstrophy are moderate, revealing that the turbulence has a spatial structure even where its intensity is weaker. The evolution equation for the total strain $s^2=s_{ij}s_{ij}$, derived from the equations of motion and shown in Eq.~(\ref{strain-prod}), contains the self-amplification term $-s_{ij}s_{jk}s_{ki}$. The corresponding self-amplification term in the evolution equation for the enstrophy $V$ reads $\omega_i\omega_js_{ij}$. Both self-amplification terms are known to be positive on average, \textit{i.e.} $\langle-s_{ij}s_{jk}s_{ki}\rangle\geq0$ and $\langle\omega_i\omega_js_{ij}\rangle\geq0$~\cite{taylor1938,tsinober1998a}.
		Moreover, these terms are found to be on average three orders of magnitude larger than the production terms due to external forcing, see brief discussion in Section \ref{Sec2}. It therefore clarifies the fundamental role played by the interaction of vorticity and strain in the dynamics of three-dimensional turbulence.\\
		The self-amplification term $\omega_i\omega_js_{ij}$ is of utmost importance in the evolution process of enstrophy. It can be expressed in terms of the eigenvalues $\Lambda_{\alpha}$ and eigenvectors ${\boldsymbol{\lambda}}_{\alpha}$ of the strain rate tensor $s_{ij}$, as
		\begin{equation}\label{eq:enstrophyproduction}
			\omega_i\omega_js_{ij}=\sum_{\alpha=1}^3\omega^2\Lambda_{\alpha}\cos^2({\boldsymbol{\omega}},{\boldsymbol{\lambda}}_{\alpha})~.
		\end{equation}
		It is also useful to decompose it as the scalar product
		\begin{equation}\label{eq:enstrophyproduction-vortexstretching}
			\omega_i\omega_js_{ij}=\omega_i(\omega_js_{ij})={\boldsymbol{\omega}}\cdot {\bf{W}}=|{\boldsymbol{\omega}}| W\cos({\boldsymbol{\omega}},{\bf{W}})~,
        \end{equation}
		where $W_i=\omega_js_{ij}$ is the $i$-th component of the vortex stretching vector ${\bf{W}}$ (with $W=|{\bf{W}}|$ and $W^2=\sum_{\alpha=1}^3\omega^2\Lambda_{\alpha}^2\cos^2({\boldsymbol{\omega}},{\boldsymbol{\lambda}}_{\alpha})$). These two expressions for the self-amplification term clearly reveal the importance of the (statistical) alignment of the vorticity vector with respect to ${\bf{W}}$ and ${\boldsymbol{\lambda}}_{\alpha}$. A well-known example of importance of the vortex stretching term ${\bf{W}}$, thus of the geometrical relation between the vector ${\boldsymbol{\omega}}$ and the eigenframe $\lbrace{\boldsymbol{\lambda}}_{\alpha}\rbrace$, is the key difference between 3D- and 2D-turbulence: for a pure 2D-flow, the eigenvectors ${\boldsymbol{\lambda}}_{\alpha}$ lie in the plane of motion, while the vector ${\boldsymbol{\omega}}$ is orthogonal to the plane, so that their scalar product vanishes. The absence of the process of vortex stretching dictates the dynamics of 2D flows.\\
		We computed the velocity gradient tensor along the trajectories of the tracer particles using a similar procedure as the one described in Ref.~\cite{luethi2007jot}. From this, we extracted the vorticity vector, the eigenvalues and the corresponding eigenvectors of the strain rate tensor. The cosine of the angle between ${\boldsymbol{\omega}}$ and ${\boldsymbol{\lambda}}_{\alpha}$ is finally computed over roughly $2000$ tracer positions and on 1 every 30 sampled time steps (thus $320$ time steps over the available $9600$), corresponding to roughly $6.4\times 10^5$ data points. Reduced sampling rates have been considered too in order to estimate the error due to different choices of the sampling step ($1.92\times 10^5$ and $6.4\times 10^4$ data points). These errors turned out to be small (in the order of one to a few percent). Fig.~\ref{fig2} shows the probability distribution functions (PDFs) of the three cosines for the reference non--rotating run.
		The data reveal a very good quantitative agreement with previously published data (see, \textit{e.g.}, refs.~\cite{kholmyansky2001pof,guala2005jfm,luethi2005jfm}). The vorticity vector reveals a strong preferential alignment with the intermediate eigenvector ${\boldsymbol{\lambda}}_2$ (the PDF of the cosine has pronounced peaks at $\pm1$), a strong statistical orthogonality with the third eigenvector ${\boldsymbol{\lambda}}_3$ (the PDF has a peak at $0$), and a weaker statistical orthogonality with the first eigenvector ${\boldsymbol{\lambda}}_1$. Despite the statistical alignment with ${\boldsymbol{\lambda}}_2$, the main contribution to the mean enstrophy production term, $\omega_i\omega_js_{ij}$, is associated with the first eigenvector ${\boldsymbol{\lambda}}_1$~\cite{kholmyansky2001pof}.
		This is explained looking at the magnitude and sign of the corresponding eigenvalues, shown in Fig.~\ref{fig4} in the following Section, which reveals that the first eigenvalue $\Lambda_1$ takes positive values only, while the second eigenvalue $\Lambda_2$ has both positive as negative values (although its average and skewness are positive). Moreover, it turns out that $\langle \Lambda_1\rangle\gg \langle\Lambda_2\rangle$.\\
		For the non-rotating case, the eigenvalues take on the values $\langle \Lambda_1\rangle=0.99\pm0.01~{\rm{s}}^{-1}$, $\langle \Lambda_2\rangle=0.11\pm0.01~{\rm{s}}^{-1}$, $\langle \Lambda_3\rangle =-1.10\pm 0.01~{\rm{s}}^{-1}$, and $\langle \Lambda_1 + \Lambda_2 + \Lambda_3\rangle\approx 0$ as expected for incompressible flows. The error margins are estimated upper bounds and are based on a comparison between the computation of PDFs with different sampling rates (sampling one time step every 30, 100 and 300 PTV time steps, respectively) and does not contain information on the measurement error (3D PTV) and those associated with computing derivatives. The ratio $\langle \Lambda_1\rangle : \langle \Lambda_2\rangle : \langle \Lambda_3\rangle$ is similar as found by L{\"{u}}thi and coworkers~\cite{luethi2005jfm} and Kholmyansky~\textit{et al.}~\cite{kholmyansky2001pof}. Further numerical details can be found in Table~\ref{tab2} in Section~\ref{Sec6}.
	\section{Geometrical statistics in the presence of background rotation}\label{Sec6}
		\begin{figure}[!t]
			\centering
			\includegraphics[width = 0.260\textwidth]{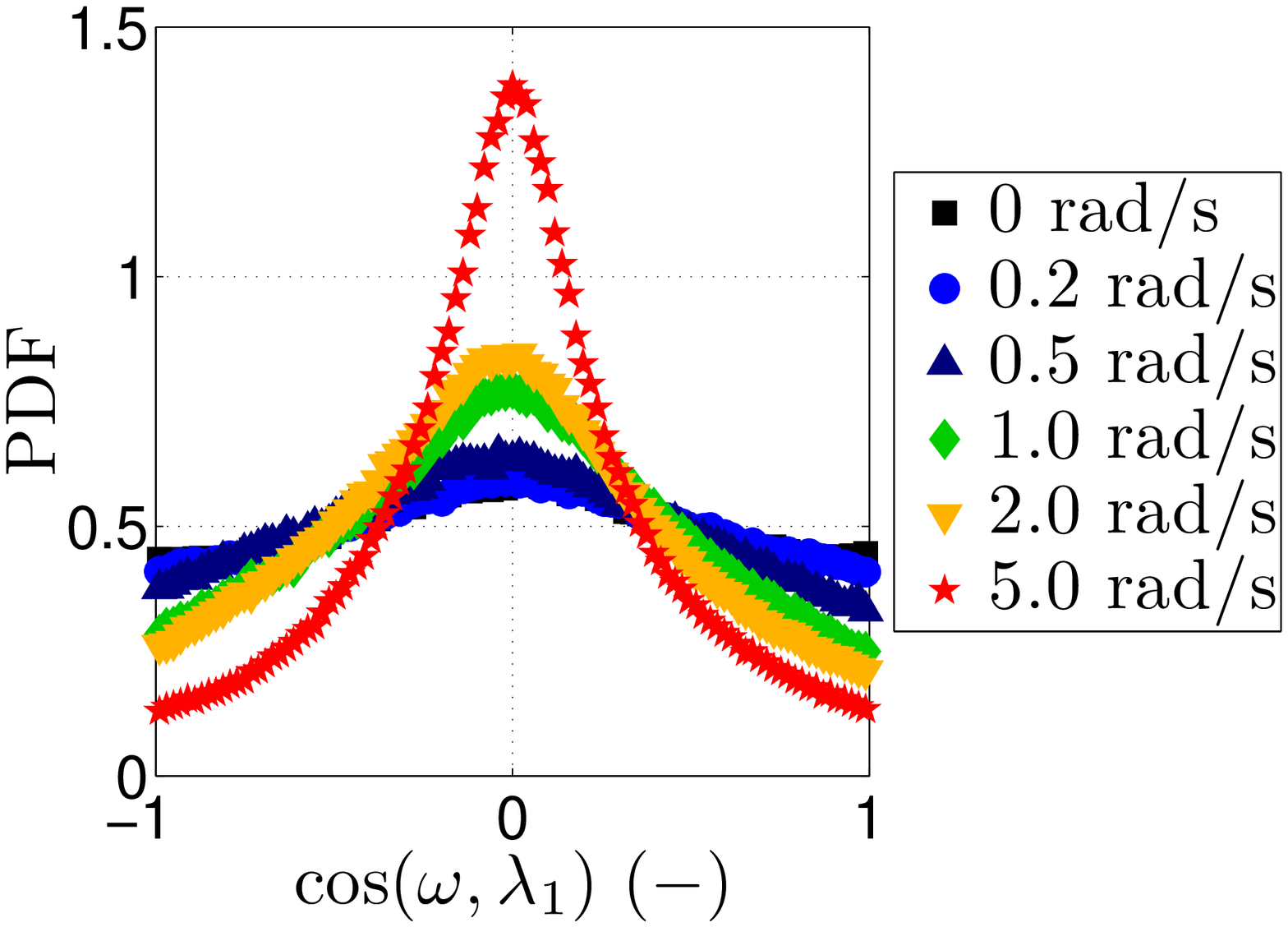}%{exps123456_cosomegalambda1-fstep30-linlin}
			\includegraphics[width = 0.245\textwidth]{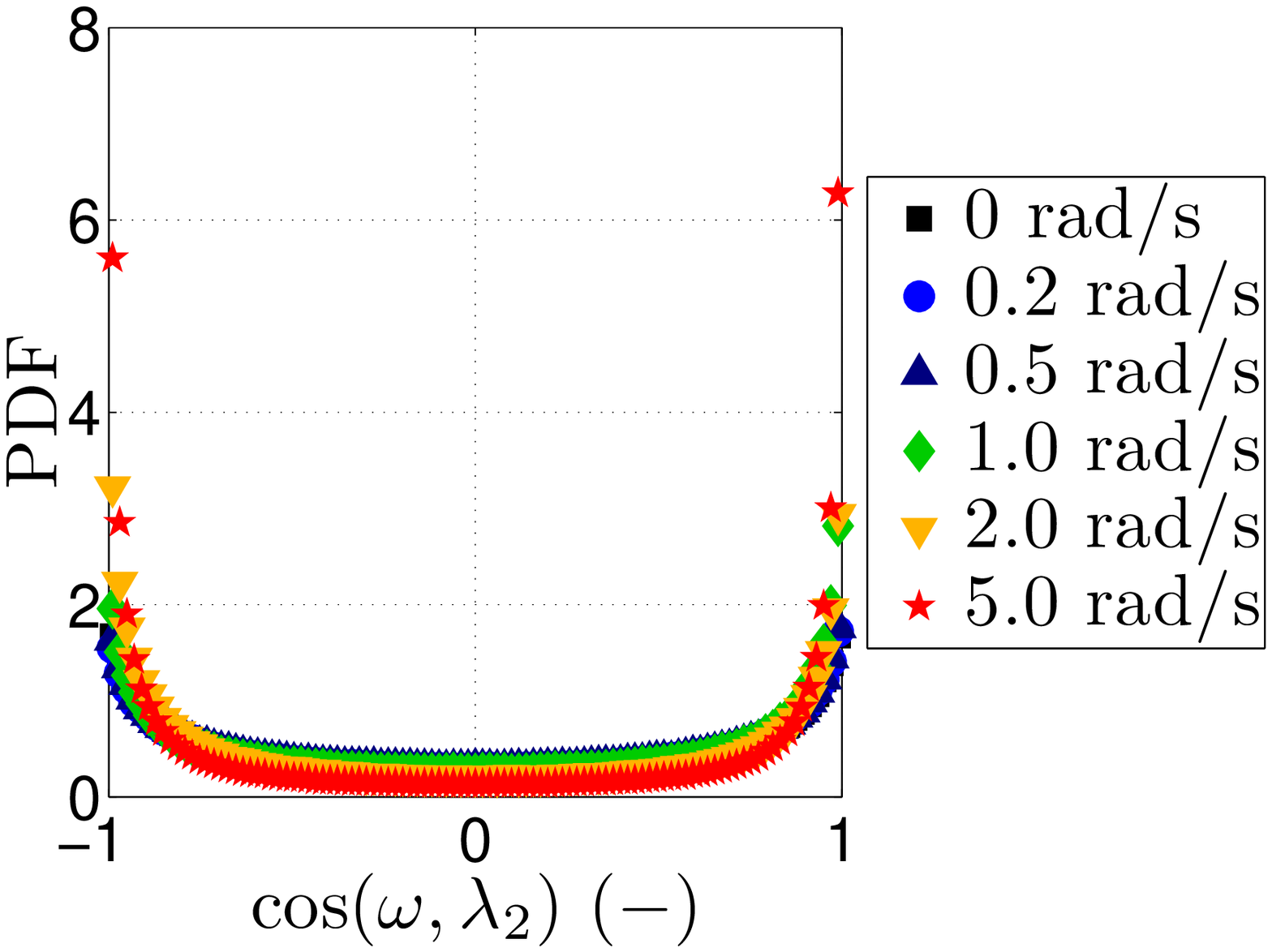}%{exps123456_cosomegalambda2-fstep30-linlin}
			\includegraphics[width = 0.260\textwidth]{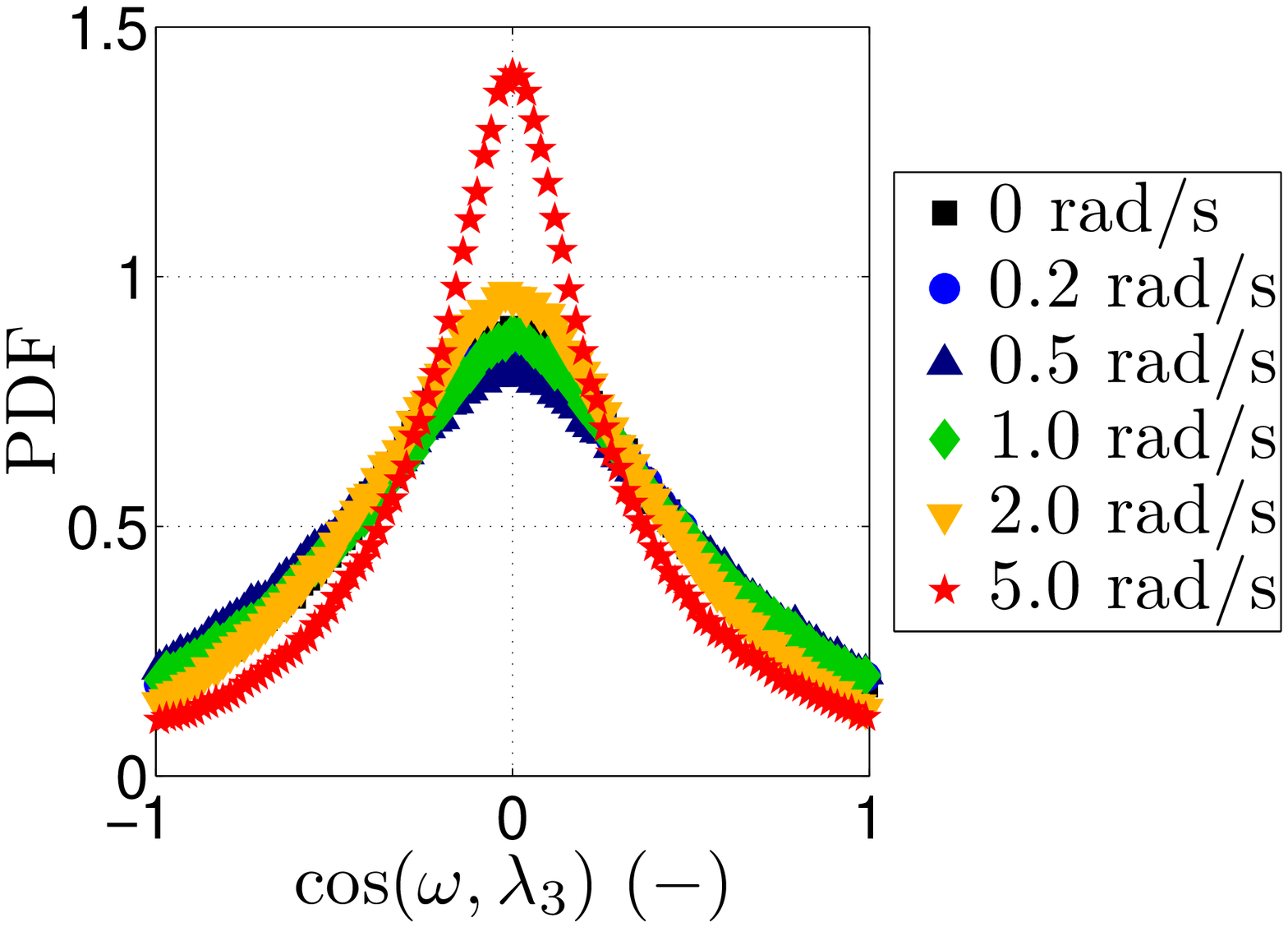}~~~~~~~~~~~~~~%{exps123456_cosomegalambda3-fstep30-linlin}
			\caption{PDFs of the cosine of the angle between the vorticity vector ${\boldsymbol{\omega}}$ and the three eigenvectors ${\boldsymbol{\lambda}}_{\alpha}$ (with $\alpha = 1, 2, 3$ in panels (a), (b), and (c), respectively) of the strain rate tensor, for all experiments with $\Omega$ varying from 0 to 5 rad/s (see legend in the right-most panel, valid also for Figs.~\ref{fig4}, \ref{fig5}, and \ref{fig6}.)}
			\label{fig3}
		\end{figure}
		The same analysis has been performed on data retrieved from the five experiments with different background rotation rates $\Omega$, and compared to the reference non-rotating run. Before discussing the results of this analysis, we present the evolution equations of the enstrophy and strain rate in the co-rotating frame of reference:
		\begin{equation}
			\frac{DV}{Dt} = \omega_i\omega_j s_{ij} + 2\Omega_i\omega_j s_{ij} + \nu \omega_i \nabla^2\omega_i + 
				\varepsilon_{ijk}\omega_i\frac{\partial F_k}{\partial x_j}~, 
			\label{vort-prod-rot}
		\end{equation}
		and
		\begin{equation}
			\frac{1}{2}\frac{Ds^2}{Dt} = -s_{ij}s_{jk}s_{ki} - \Omega_i\omega_j s_{ij} - \frac{1}{4}\omega_i\omega_j s_{ij} - 
				s_{ij}\frac{1}{\rho}\frac{\partial^2 p}{\partial x_i\partial x_j} + \nu s_{ij}\nabla^2 s_{ij} + s_{ij}F_{ij}~.
			\label{strain-prod-rot}
		\end{equation}
		The corrections due to background rotation for both evolution equations are of similar form, and reduce to the simplified expression $\Omega \omega_j s_{3j}=\Omega({\boldsymbol{\omega}}\cdot {\boldsymbol{\nabla}})w$ for $\Omega_3 = |{\bf{\Omega}}| = \Omega$ and $w$ the vertical velocity component (parallel to the rotation axis).\\
		The PDFs of the cosine of the angle between ${\boldsymbol{\omega}}$ and each ${\boldsymbol{\lambda}}_{\alpha}$ are shown in the three panels of Fig.~\ref{fig3}. A first observation is that the PDFs remain symmetric with increasing rotation rate. This is supported by the computed skewness for each PDF, which always remains very small. 
		While the PDFs show only minor differences when the rotation rate is increased from $0$ to $0.2$ and up to $0.5~\mathrm{rad/s}$, a background rotation $\Omega\in\lbrace1.0;2.0\rbrace~\mathrm{rad/s}$, and an even faster rotation $\Omega=5.0~\mathrm{rad/s}$, are seen to significantly affect the geometrical dynamical structure of the flow field. The PDFs for $\Omega=1.0$ and $\Omega=2.0~\mathrm{rad/s}$ have similar shapes, while the effects of rotation get substantially more pronounced at $5.0~\mathrm{rad/s}$. The changes are monotonic with increasing $\Omega$. For fast rotation rates, the vorticity vector is seen to be significantly more aligned -- in a statistical sense -- with the second eigenvector ${\boldsymbol{\lambda}}_2$, and almost perpendicular to the other two eigenvectors.\\
		\begin{figure}[!b] % EIGENVALUES, rotating
			\centering
			\includegraphics[width = 0.32\textwidth]{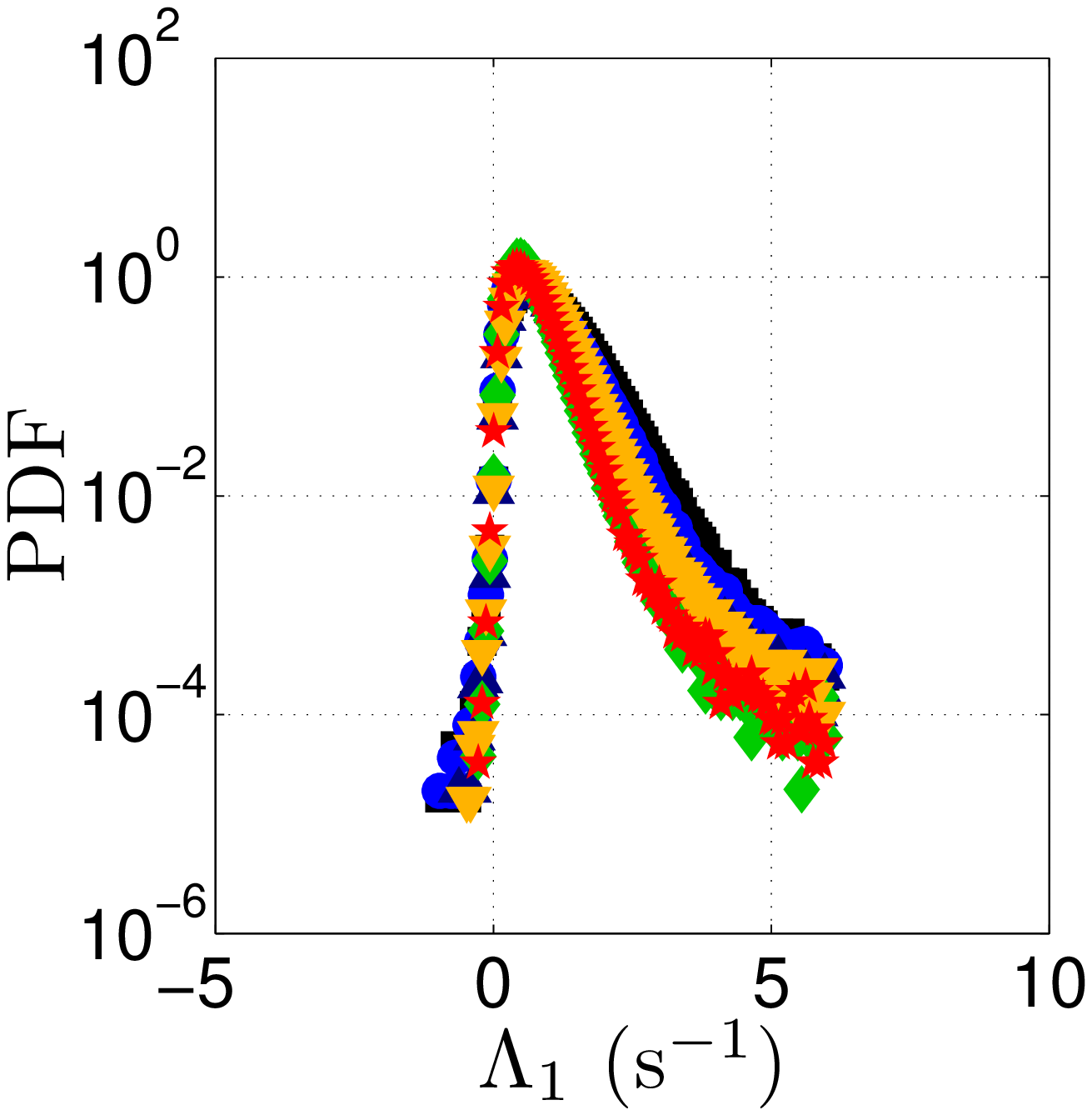}%{exps123456_lambdaval1-fstep30-linlog}
			\includegraphics[width = 0.32\textwidth]{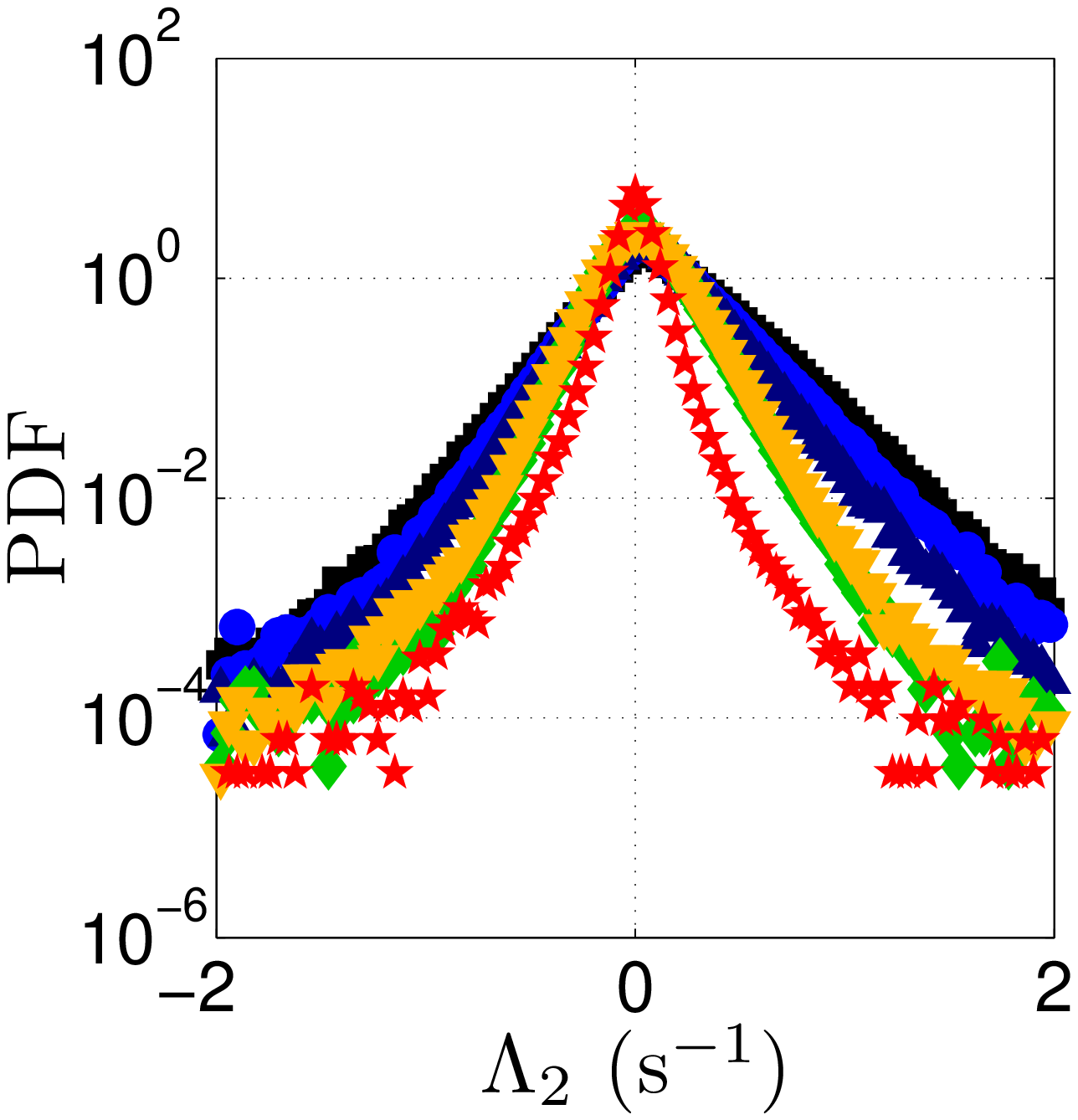}%{exps123456_lambdaval2-fstep30-linlog}
			\includegraphics[width = 0.32\textwidth]{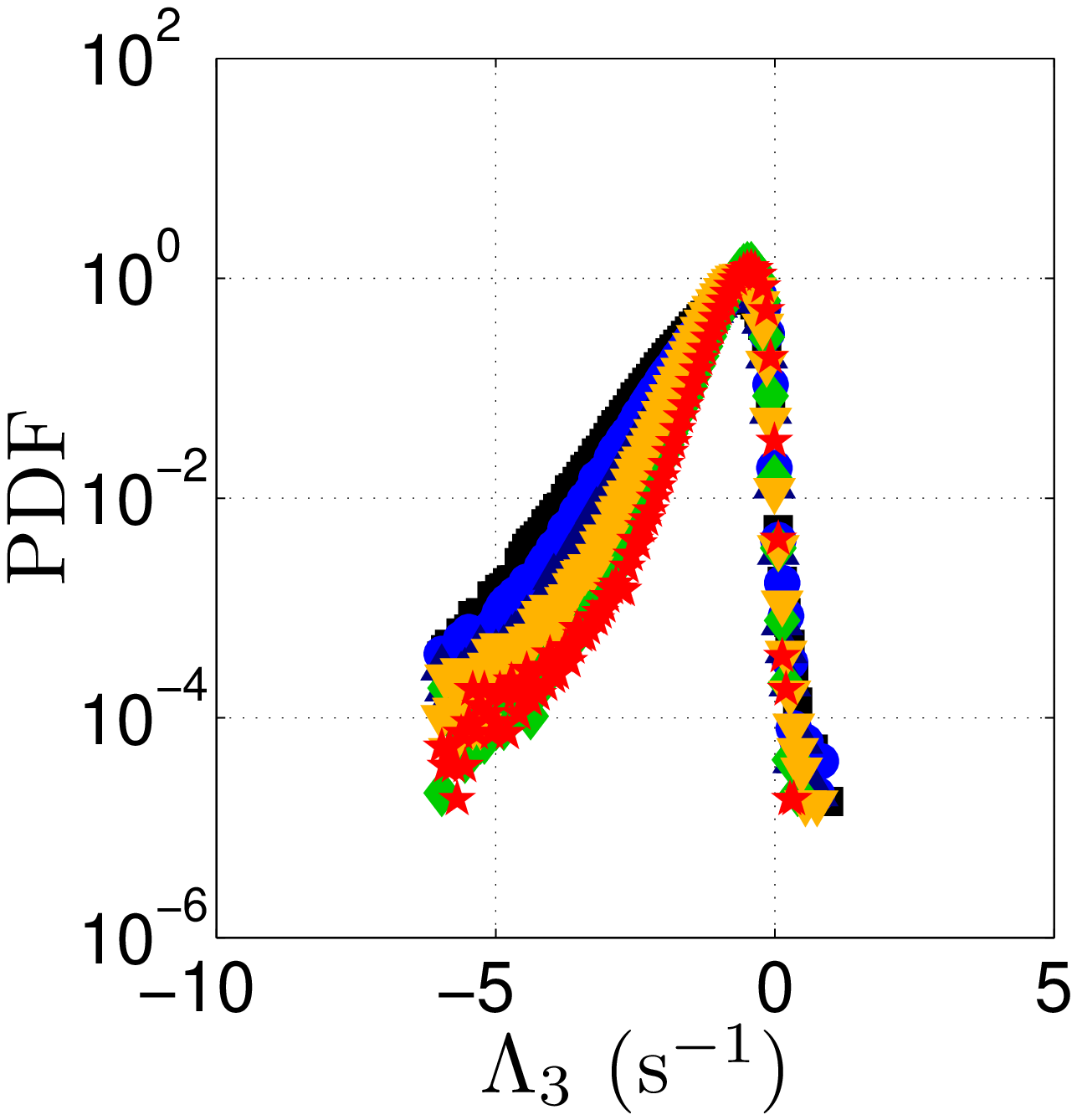}%{exps123456_lambdaval3-fstep30-linlog}
			\caption{PDFs of the three eigenvalues $\Lambda_{\alpha}$ (with $\alpha = 1, 2, 3$ in panels (a), (b), and (c), respectively) of the strain rate tensor, for all experiments with $\Omega$ varying from 0 to 5 rad/s. Symbols as in Fig. \ref{fig3}.}
			\label{fig4}
		\end{figure}
		\linespread{1.3}
		\begin{table}[!b]
			\begin{center}
				\begin{tabular}{lccccccc}
					\hline
					$\Omega$ \footnotesize{(rad/s)} & & 0 & 0.2 & 0.5 & 1.0 & 2.0 & 5.0\\
					\hline
					\hline
					$\langle \Lambda_1\rangle$ \footnotesize{($\mathrm{s}^{-1}$)}
						&  & 0.993 & 0.853 & 0.804 & 0.653 & 0.805 & 0.642 \\
					$\langle \Lambda_2\rangle$ \footnotesize{($\mathrm{s}^{-1}$)}
						&  & 0.113 & 0.090 & 0.072 & 0.037 & 0.035 & 0.002 \\
					$\langle \Lambda_3\rangle$ \footnotesize{($\mathrm{s}^{-1}$)}
						&  & -1.101 & -0.935 & -0.871 & -0.685 & -0.839 & -0.646 \\
					$\langle V\rangle^{1/2}$ \footnotesize{($\mathrm{s}^{-1}$)}
						&  & 1.29 & 1.17 & 1.15 & 0.98 & 1.13 & 0.92 \\
					$\langle s^2\rangle^{1/2}$ \footnotesize{($\mathrm{s}^{-1}$)}
						&  & 1.60 & 1.42 & 1.34 & 1.08 & 1.30 & 1.05 \\
					\hline
					$\langle \Lambda_1 + \Lambda_2 + \Lambda_3\rangle$ \footnotesize{($\mathrm{s}^{-1}$)}
						&  & 0.005 & 0.008 & 0.005 & 0.005 & 0.001 & -0.002 \\
					$\langle \Lambda_1^2 + \Lambda_2^2 + \Lambda_3^2\rangle/\langle s^2\rangle$ \footnotesize{(--)}
						&  & 1.26 & 1.18 & 1.10 & 1.04 & 1.07 & 1.03 \\
					\hline
					$\sigma_{\Lambda_2}$ \footnotesize{($\mathrm{s}^{-1}$)}
						&  & 0.35 & 0.30 & 0.26 & 0.19 & 0.20 & 0.10 \\
					$S_{\Lambda_2}$ \footnotesize{(--)}
						&  & 1.3 & 1.3 & 1.1 & 0.8 & 0.8 & 0.1 \\
					\hline\\
				\end{tabular}
				\linespread{1}
				\caption{For each (non) rotating experiment, we have summarised the averaged values of the eigenvalues $\Lambda_{\alpha}$, the average enstrophy $\langle V\rangle^{1/2} = \langle \frac{1}{2} \omega^2\rangle^{1/2}$, the average strain $\langle s^2\rangle^{1/2}$ (with $s^2=s_{ij}s_{ij}$), the sum $\langle\sum_{\alpha=1}^3 \Lambda_{\alpha}\rangle$ (which should be zero for incompressible flows), the sum $\langle\sum_{\alpha=1}^3 \Lambda_{\alpha}^2\rangle/\langle s^2\rangle$ (which should by definition be equal to unity), and the root-mean-square value $\sigma$ and skewness $S$ of the intermediate eigenvalue $\Lambda_2$, respectively, as function of the rotation rate. Both $\sigma_{\Lambda_1}$ and $\sigma_{\Lambda_3}$ decrease from approximately $1.3~{\rm{s}}^{-1}$to $0.7~{\rm{s}}^{-1}$ when the rotation rate is increased from 0 to 5 rad/s. For $\Lambda_1$ and $\Lambda_3$ we find for the skewness: $S_{\Lambda_1}\approx 1.5$ and $S_{\Lambda_3}\approx -1.5$, almost independent of the rotation rate $\Omega$.}
				\label{tab2}
			\end{center}
		\end{table}\linespread{1}
		Figure~\ref{fig4} shows the PDFs of the eigenvalues $\Lambda_{\alpha}$ associated with the eigenvectors ${\boldsymbol{\lambda}}_{\alpha}$ of the strain rate tensor. A clear effect of the system rotation is visible, particularly on the PDF of $\Lambda_2$. Without rotation, the PDF of $\Lambda_2$ is slightly asymmetric with positive skewness. However, the skewness gradually decreases with increasing rotation rate and the PDFs of $\Lambda_2$ become more symmetric around zero, which is nicely reflected by $\langle \Lambda_2\rangle \approx 0$ for $\Omega = 5~{\rm{rad/s}}$ (and absence of any appreciable skewness). The observation that for large rotation rate $\langle \Lambda_2\rangle \rightarrow 0$ and $\langle \Lambda_1\rangle \approx -\langle \Lambda_3\rangle$ is consistent with growing two-dimensionalisation of the flow. The main effect of increasing rotation rate on the distributions of $\Lambda_1$ and $\Lambda_3$ is the suppression of the extreme values of the largest and smallest eigenvalues.
		Mean value, root-mean-square, and skewness of the eigenvalues $\Lambda_{\alpha}$ obtained for the experiments with different rotation rates are reported in Table~\ref{tab2}.\\
		The evolution of the local enstrophy and strain is governed by the enstrophy and strain self-amplification terms, $\omega_i\omega_js_{ij}$ and $-s_{ij}s_{jk}s_{ki}$, respectively, see Eqs.~(\ref{vort-prod-rot}) and (\ref{strain-prod-rot}). Before analysing the effect of rotation on these self-amplification terms, we briefly have a look at enstrophy and strain. From Table~\ref{tab2} we can conclude that indeed $\Sigma_{\alpha=1}^3 \langle \Lambda_{\alpha}^2\rangle \approx \langle s^2\rangle$, as should by definition be the case. Note that the difference between $\langle \Lambda_1^2 + \Lambda_2^2 + \Lambda_3^2\rangle$ and $\langle s^2\rangle$ is due to lack of mass conservation when extracting the velocity derivatives from our 3D-PTV data (with the number of particles currently used for tracking one can hardly get better results).
		Both the rms values of the strain and the enstrophy tend to decrease with increasing rotation rate, which is predominantly due to a strong reduction of extreme events for local strain and enstrophy. This reduction is basically monotonous with increasing rotation rate, with an exception for $\Omega = 2$ rad/s, see Table~\ref{tab2}.
		For homogeneous and incompressible turbulent flows one would expect the following equality: $\langle s^2\rangle = \langle V\rangle$. As we cannot claim to satisfy homogeneity conditions (in particular in the vertical direction homogeneity is absent) the ratio $\langle s^2\rangle/\langle V\rangle$ indeed tends to approach one (from above) with increasing rotation rate. This is consistent with enhanced homogeneity in the vertical direction when system rotation increases, see Ref. \cite{bokhoven2009pof}.
		The vertical homogeneity of the turbulent kinetic energy (and several other turbulence-related quantities) for $\Omega\geq1.0~\mathrm{rad/s}$ let us conclude that within the measurement volume there is no detectable spatial transition from 3D to Q2D flow dynamics for such rotation rates. Actually, the turbulence in the measurement volume is found to be fully rotation dominated, therefore we are not averaging in space a 3D state with a Q2D one. The results for $\Omega=0$ are fully consistent with the literature, despite the flow is vertically inhomogeneous. We expect the mild rotation rate $\Omega=0.2~\mathrm{rad/s}$ to behave similarly and is hardly affected by the system rotation.
		It is instead possible that a transition between two states is present within the measurement volume at a certain height $z_*$ for $\Omega=0.5~\mathrm{rad/s}$, despite our data for this run do not show anomalous features. Nevertheless, our statistical results show a gradual shift from the non-rotating case -- benchmarked with the literature -- to the fastest rotating one, apart from the aforementioned $\Omega=2.0~\mathrm{rad/s}$ run (which is not related to a possible transition from 3D to Q2D flow behavior).\\
		In Section~\ref{Sec5}, the self-amplification term $\omega_i\omega_js_{ij}$ of the evolution equation of the enstrophy, see Eq.~(\ref{vort-prod}), was expressed in terms of the inner product of the vorticity vector ${\boldsymbol{\omega}}$ and the stretching vector ${\bf{W}}$, see Eq.~(\ref{eq:enstrophyproduction-vortexstretching}). L{\"{u}}thi and coworkers~\cite{luethi2005jfm} gave a physical interpretation of the geometrical invariant $\cos({\boldsymbol{\omega}},{\bf{W}})$ by analysing and comparing both unconditioned and conditioned (on $\omega^2$) PDFs of $\cos({\boldsymbol{\omega}},{\bf{W}})$. For $\cos({\boldsymbol{\omega}},{\bf{W}})>0$ vortex stretching occurs, while $\cos({\boldsymbol{\omega}},{\bf{W}})<0$ indicates vortex compression (for $\cos({\boldsymbol{\omega}},{\bf{W}})=0$ only tilting occurs).
		The PDFs conditioned on the enstrophy revealed that $\langle \omega_i\omega_js_{ij}\rangle>0$, see Fig.~9c in Ref.~\cite{luethi2005jfm} and Fig.~4 in Ref.~\cite{kholmyansky2001pof}, thus supporting the conjecture by L{\"{u}}thi and coworkers~\cite{luethi2005jfm} that the strong positive skewness of the PDF of $\cos({\boldsymbol{\omega}},{\bf{W}})$ indeed indicates positiveness of the mean enstrophy production. In Fig. \ref{fig5}(a) we have shown the PDF of $\cos({\boldsymbol{\omega}},{\bf{W}})$ for the (non) rotating experiments. Increasing rotation rate first reduces vortex stretching events and enhances vortex compression events (for rotation rates up to $1~\mathrm{rad/s}$). For higher rotation rates, also vortex compression becomes less probable, and for the highest rotation rate in our experiments the PDF of $\cos({\boldsymbol{\omega}},{\bf{W}})$ has become almost symmetric around $\cos({\boldsymbol{\omega}},{\bf{W}})=0$ and is dominated by tilting events.
		Vortex stretching and compression events become rare, which is another strong indicator for two-dimensionalisation of the turbulent flow. Such results are in good agreement with previously published data~\cite{kinzel2010}, at least for the slowest rotating runs which anticipate the trend.\\
		\begin{figure}[!t]
			\centering
			\includegraphics[width = 0.32\textwidth]{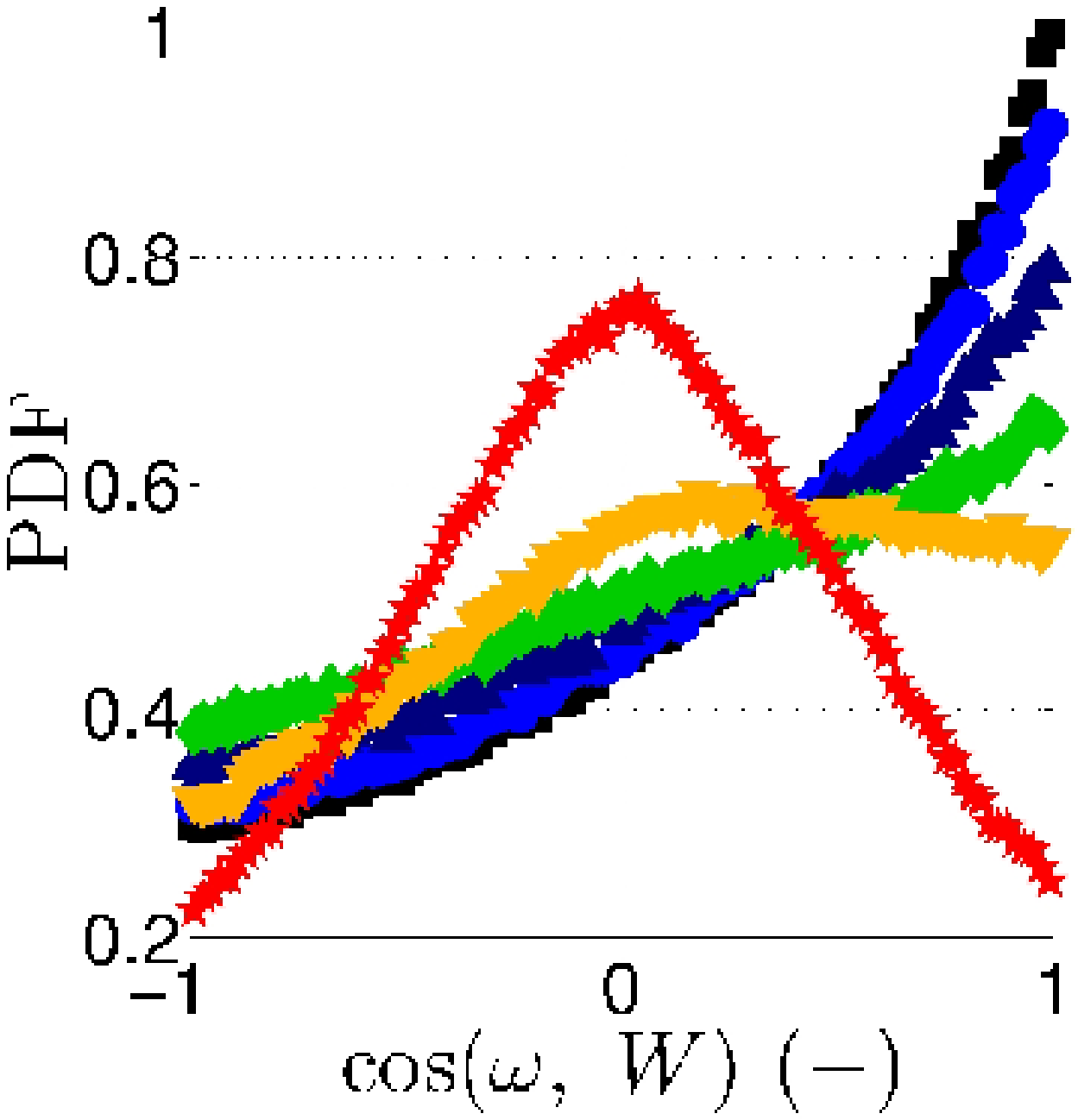}%{exps123456_cosomegavstretch-fstep30-linlin}
			\includegraphics[width = 0.33\textwidth]{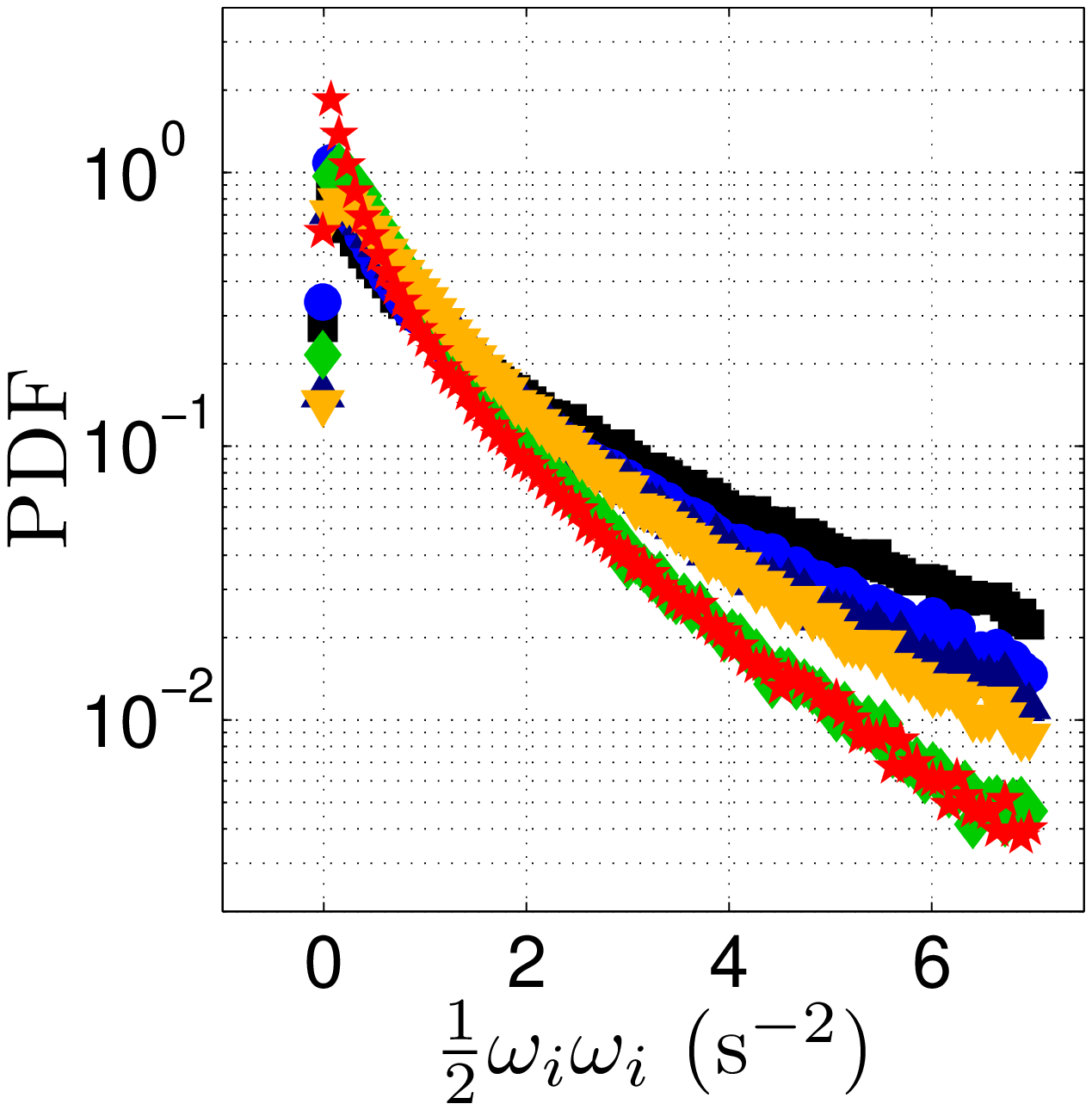}%{exps123456_enstrophy-fstep100-linlog}
			\includegraphics[width = 0.33\textwidth]{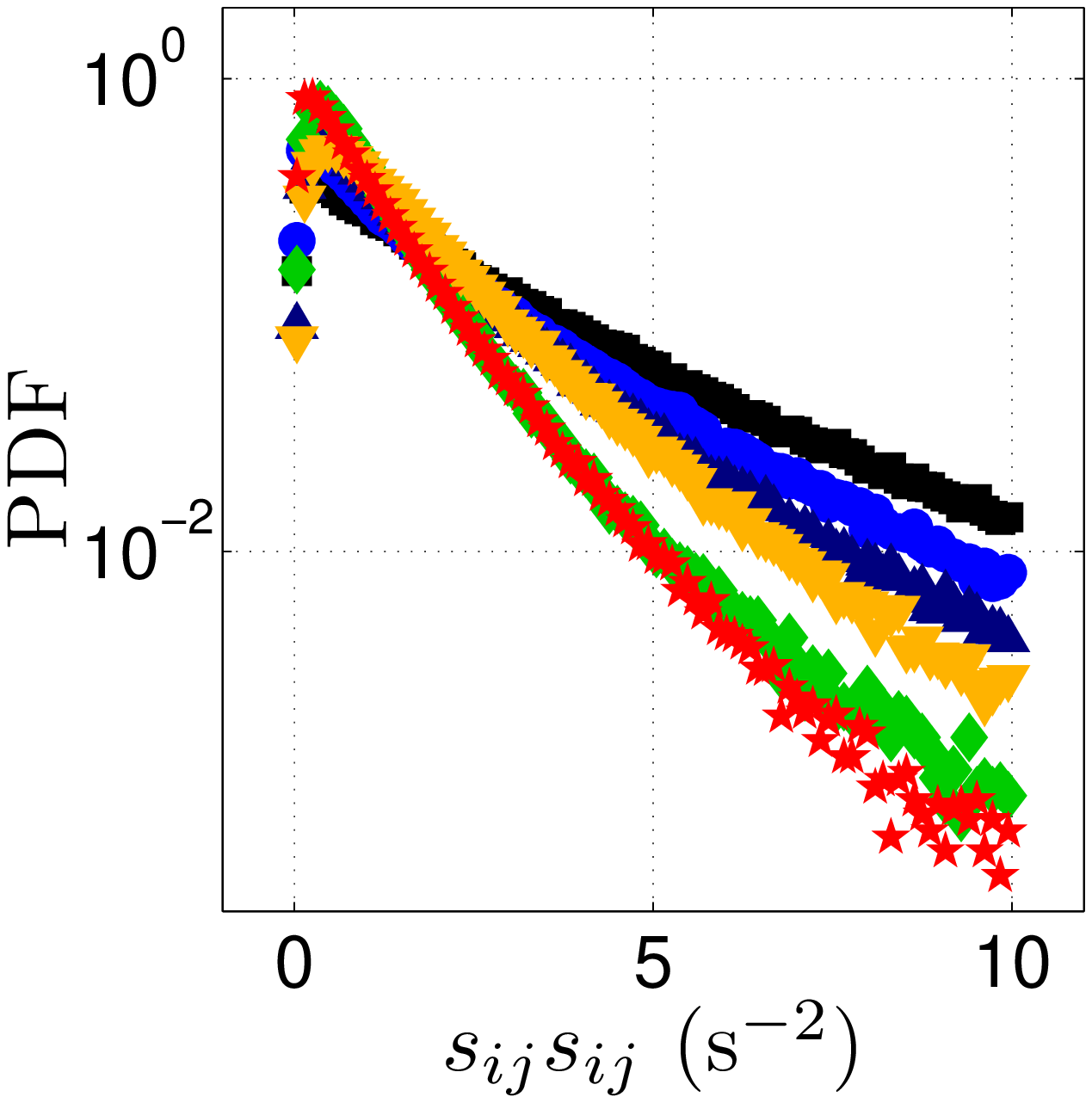}\\%{exps123456_strainsquared-fstep100-linlog}
			\includegraphics[width = 0.33\textwidth]{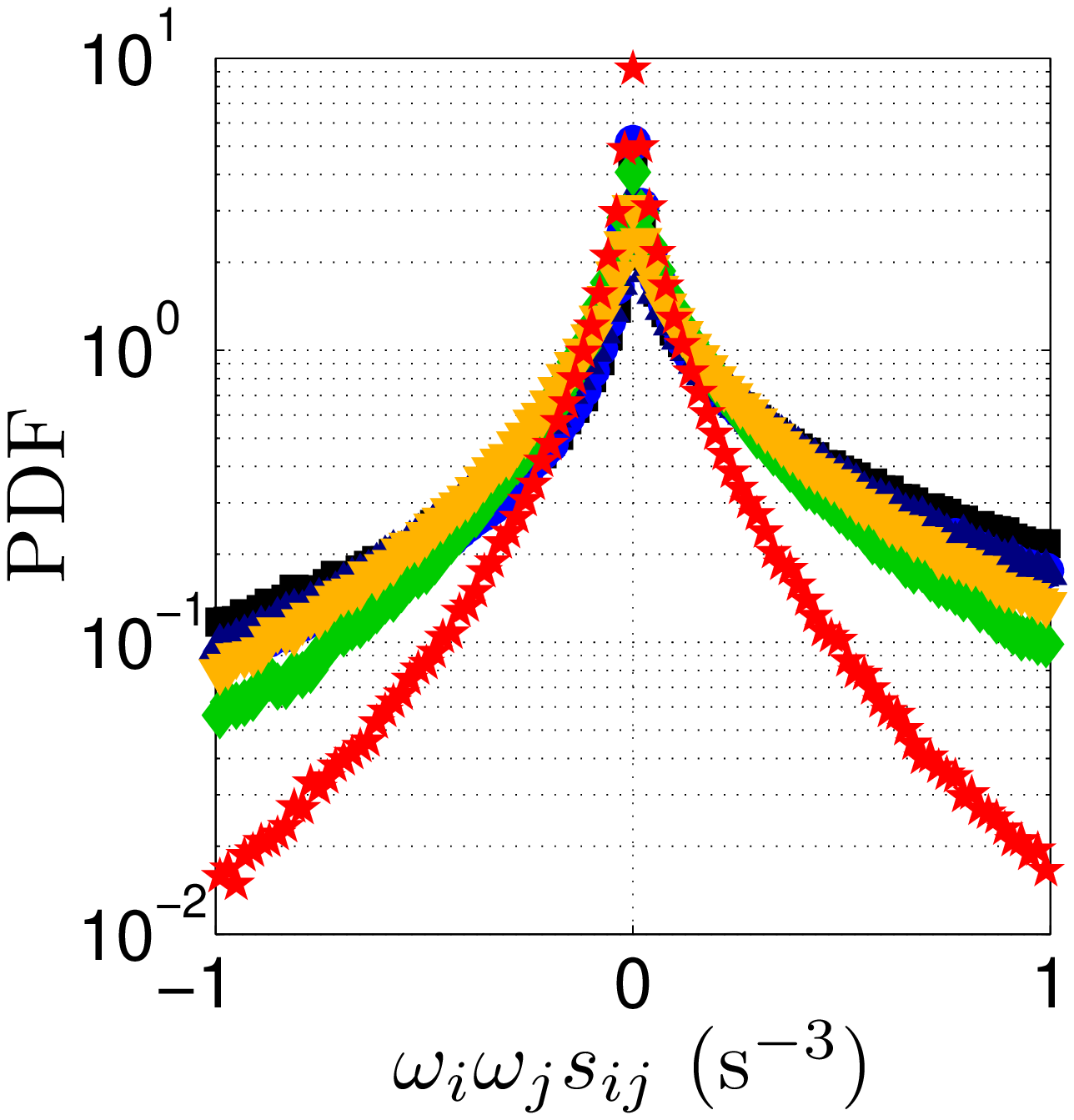}%{exps123456_enstrprod-fstep30-linlog}
			\includegraphics[width = 0.33\textwidth]{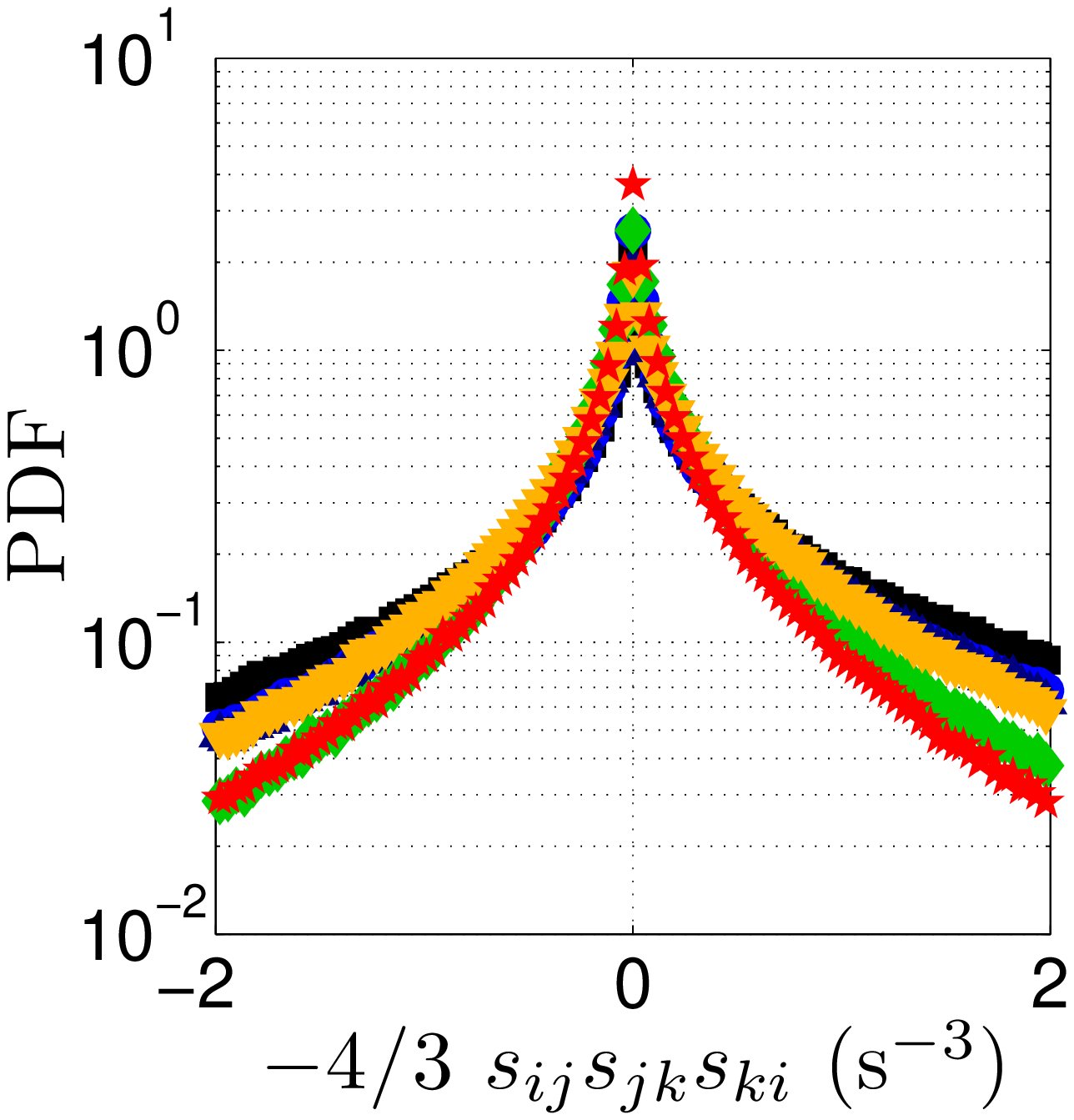}%{exps123456_strainprod-fstep30-linlog}
			\caption{Panel (a): PDFs of the geometrical invariant $\cos({\boldsymbol{\omega}},{\bf{W}})$, with ${\bf{W}}$ the vortex stretching vector, for the different rotation rates. Panels (b) and (c): PDFs of the enstrophy $\frac{1}{2}\omega_i\omega_i$ and of the squared strain rate $s_{ij}s_{ij}$ for the different rotation rates. Panels (d) and (e): PDFs of the enstrophy production contribution $\omega_i\omega_js_{ij}$ and the strain production term $-\frac{4}{3}s_{ij}s_{jk}s_{ki}$ for all experiments with different rotation rates (with $\Omega$ varying from 0 to 5 rad/s). Symbols as in Fig.~\ref{fig3}.}
			\label{fig5}
		\end{figure}
		The effects of system rotation on the enstrophy, on the squared strain rate, and on the self-amplification terms of enstrophy and strain are also illustrated in Fig.~\ref{fig5}: in Fig.~\ref{fig5}(b) we show the enstrophy distribution function, which right tail gets considerably lower with increasing rotation rate; the same effect is observed on the distribution function of the squared strain rate, shown in Fig.~\ref{fig5}(c). Fig.~\ref{fig5}(d) displays the PDF of the enstrophy self-amplification term $\omega_i\omega_js_{ij}$. The PDF of the strain self-amplification term $-\frac{4}{3}s_{ij}s_{jk}s_{ki}$ is shown in Fig.~\ref{fig5}(e). Both PDFs are slightly skewed for the non-rotating case, but the skewness weakens with increasing rotation rate and almost fully disappears for the highest rotation rate.
		The tails of the PDFs tend to decrease with increasing rotation rate: production of extreme values of enstrophy or strain is less probable in rotating turbulence (which is also reflected in the lack of extreme events in the PDFs of enstrophy and strain). Additionally, we observe that the tails of the PDF of $\omega_i\omega_js_{ij}$ reduce even considerably for the highest rotation rate, clearly indicating the lack of enstrophy production in rapidly rotating turbulence.
		This is fully in agreement with the two-dimensionalisation mechanism, as enstrophy production is virtually absent for quasi-two-dimensional turbulence. This is confirmed by the quantitative values of $P_e=\langle\omega_i\omega_js_{ij}\rangle$ with increasing rotation rate (see Table~\ref{tab3}) which become very small (for $\Omega = 5~{\rm{rad/s}}$) when compared to the non-rotating value. An overview of the numerical data for the enstrophy and strain self-amplification terms and some relevant statistical quantities are summarised in Table~\ref{tab3}.
		\linespread{1.3}
		\begin{table}[!b]
			\begin{center}
				\begin{tabular}{lccccccc}
					\hline
					$\Omega$ \footnotesize{(rad/s)} & & 0 & 0.2 & 0.5 & 1.0 & 2.0 & 5.0\\
					\hline
					\hline
					$P_e=\langle\omega_i\omega_js_{ij}\rangle$ \footnotesize{($\mathrm{s}^{-3}$)}
						&  & 0.07 & 0.06 & 0.05 & 0.04 & 0.04 & 0.00 \\
					$S_{P_e}$ \footnotesize{(--)}
						&  & 0.5 & 0.5 & 0.4 & 0.5 & 0.4 & 0.1 \\
					\hline
					$P_s=-\frac{4}{3}\langle s_{ij}s_{jk}s_{ki}\rangle$ \footnotesize{($\mathrm{s}^{-3}$)}
						&  & 0.06 & 0.05 & 0.05 & 0.03 & 0.04 & 0.00 \\
					$S_{P_s}$ \footnotesize{(--)}
						&  & 0.2 & 0.2 & 0.3 & 0.3 & 0.2 & 0.0 \\
					\hline\\
				\end{tabular}
				\linespread{1}
				\caption{For each (non)rotating experiment, we have summarised the averaged values of the enstrophy and strain self-amplification terms, $P_e=\langle \omega_i\omega_js_{ij}\rangle$ and $P_s=-\frac{4}{3}\langle s_{ij}s_{jk}s_{ki}\rangle$, respectively, and the skewnesses $S_{P_e}$ and $S_{P_s}$.}
				\label{tab3}
			\end{center}
		\end{table}\linespread{1}
		From the data we can also conclude that $P_e \approx P_s$ (for homogeneous and incompressible flows $P_e=P_s$), with $P_s=-\frac{4}{3}\langle s_{ij}s_{jk}s_{ki}\rangle$, and show the same trend for increasing rotation rate. The relatively small differences for zero or small rotation rates are most likely due to the absence of (vertical) homogeneity. For larger rotation rates, vertical homogeneity is more or less restored. Note that the pointwise sum of the contributions of $-\frac{4}{3}s_{ij}s_{jk}s_{ki}$ and $\omega_i\omega_j s_{ij}$ do not vanish (data not shown here).
		The external contribution to the enstrophy and strain rate of change due to system rotation, denoted by $P_{\Omega}=\langle \Omega_i\omega_js_{ij}\rangle$, is found to be several orders of magnitude smaller than $P_e$ and $P_s$. We found $P_{\Omega}\approx 10^{-3}~{\rm{s}}^{-3}$ for weak rotation rates ($\Omega \lesssim 0.5$ rad/s) and $P_{\Omega}\approx 10^{-4}~{\rm{s}}^{-3}$ for the highest ones, thus supporting our assertion that enstrophy and strain production is largely governed by the contributions $\omega_i\omega_j s_{ij}$ and $s_{ij}s_{jk}s_{ki}$, respectively.
		Furthermore, the tails of the (almost symmetric) distribution of the term $\Omega_i\omega_js_{ij}$ (not shown here) are seen to get monotonically lower as the rotation rate is increased -- as expected: the term vanishes in the case of pure 2D dynamics.\\
		The PDFs shown in Figs.~\ref{fig4} and \ref{fig5} and the data in Table~\ref{tab2}, clearly show that the rotation rate $\Omega=2~{\rm{rad/s}}$ represents a special case. Like in our studies of the Lagrangian velocity and acceleration PDFs and autocorrelations~\cite{delcastello2011pre,delcastello2011prl} it turns out that this particular rotation rate is close to a critical Rossby number ${\rm{Ro}}\approx 0.1$, see brief discussion in Section~\ref{Sec4} and in Refs.~\cite{delcastello2011pre,delcastello2011prl}, which apparently contributes to enhanced extreme events affecting rms values of, e.g., enstrophy and strain.\\
		\begin{figure}[!t]
			\centering
			\includegraphics[height = 0.205\textheight]{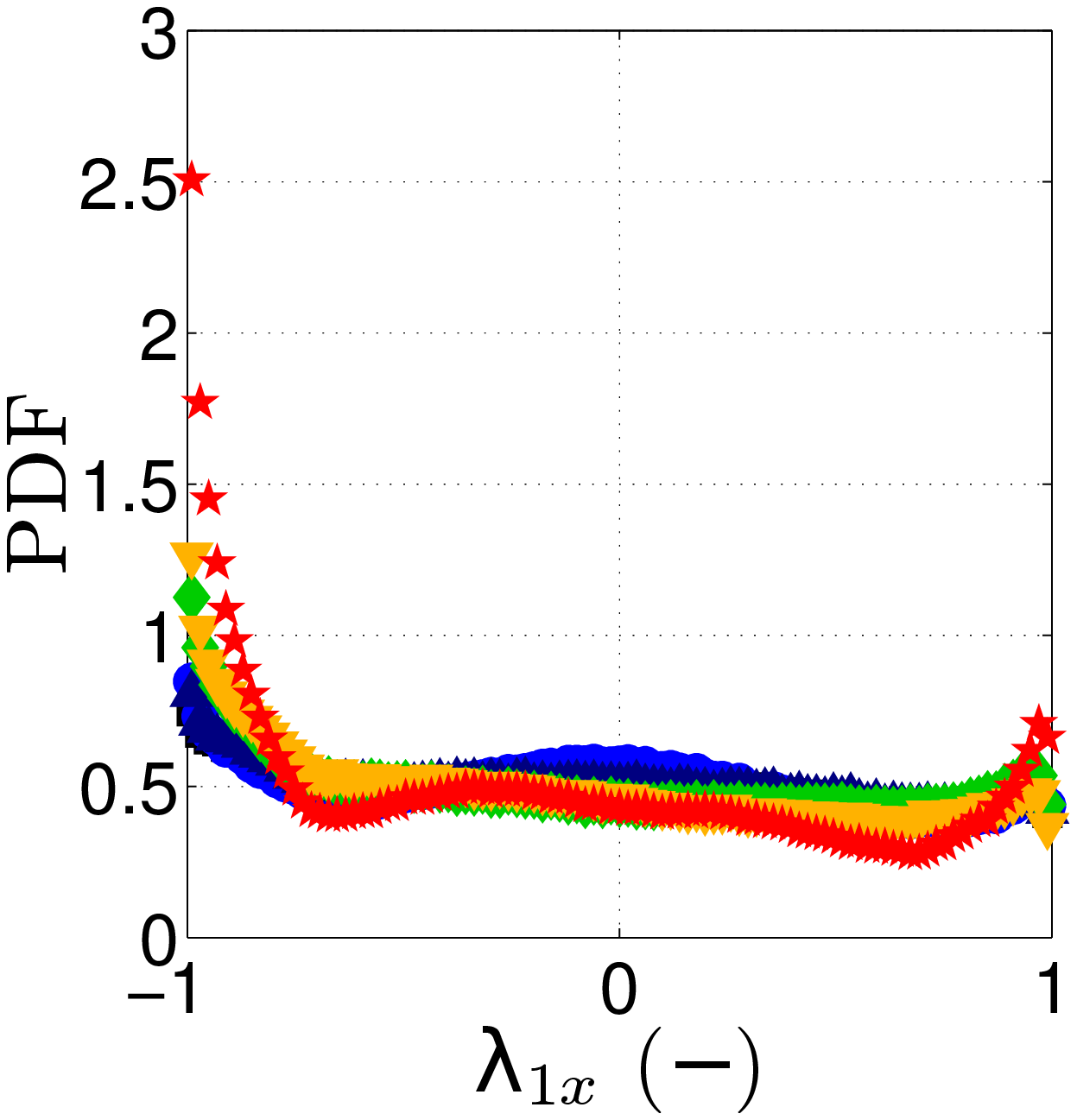}~%{exps123456_lambda1x-fstep30-linlin}
			\includegraphics[height = 0.205\textheight]{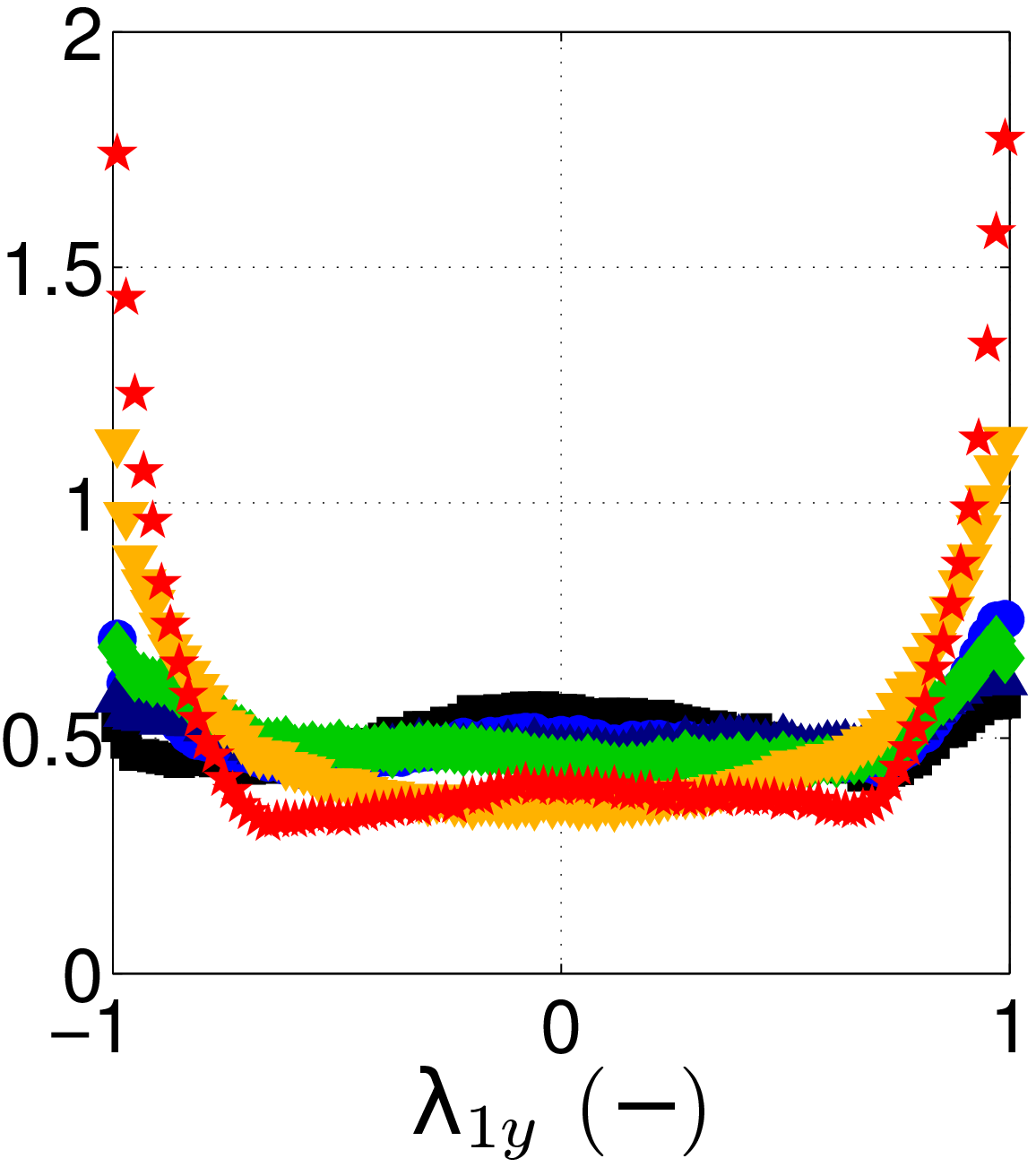}~%{exps123456_lambda1y-fstep30-linlin}
			\includegraphics[height = 0.205\textheight]{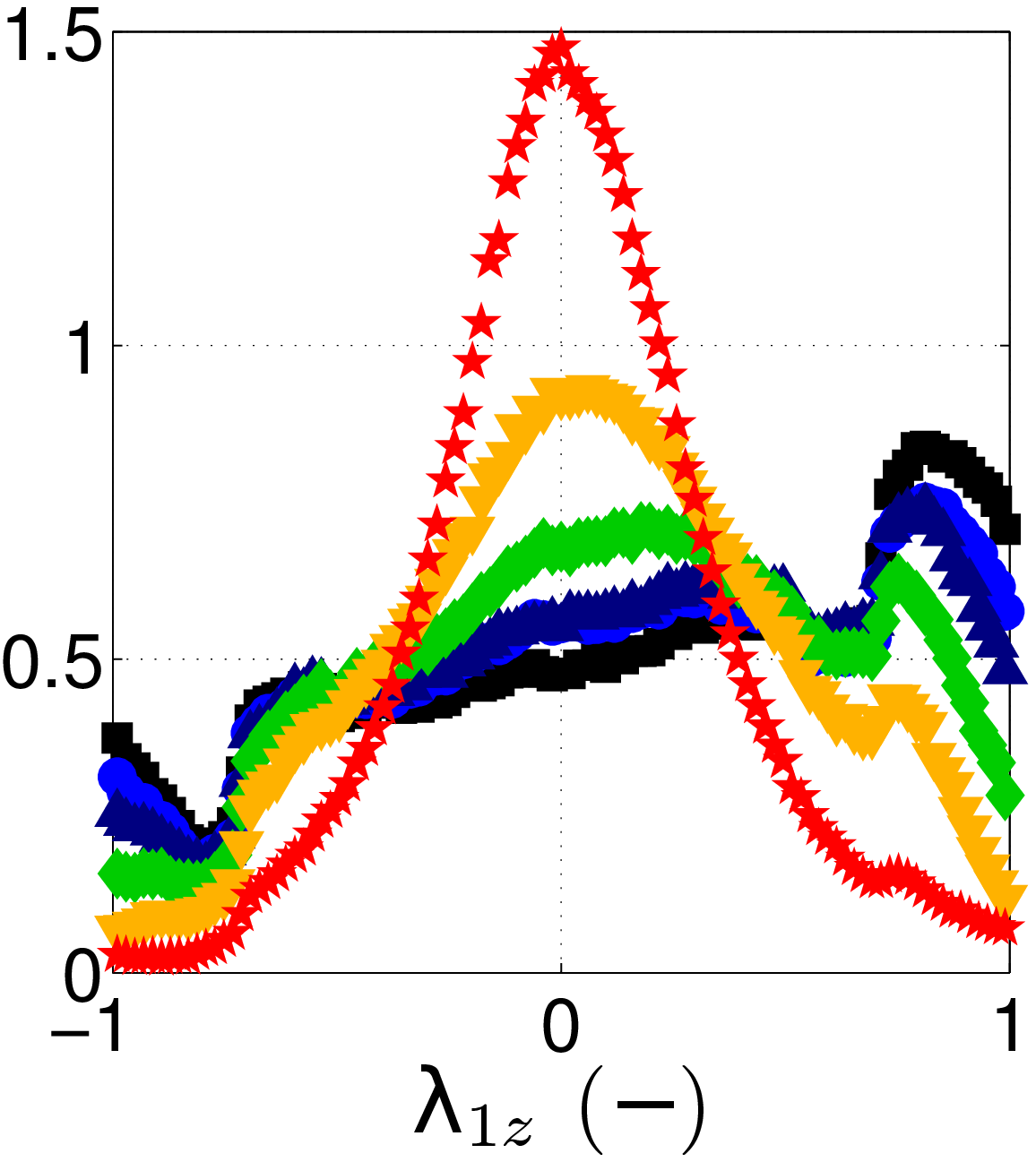}\\%{exps123456_lambda1z-fstep30-linlin}
			\includegraphics[height = 0.205\textheight]{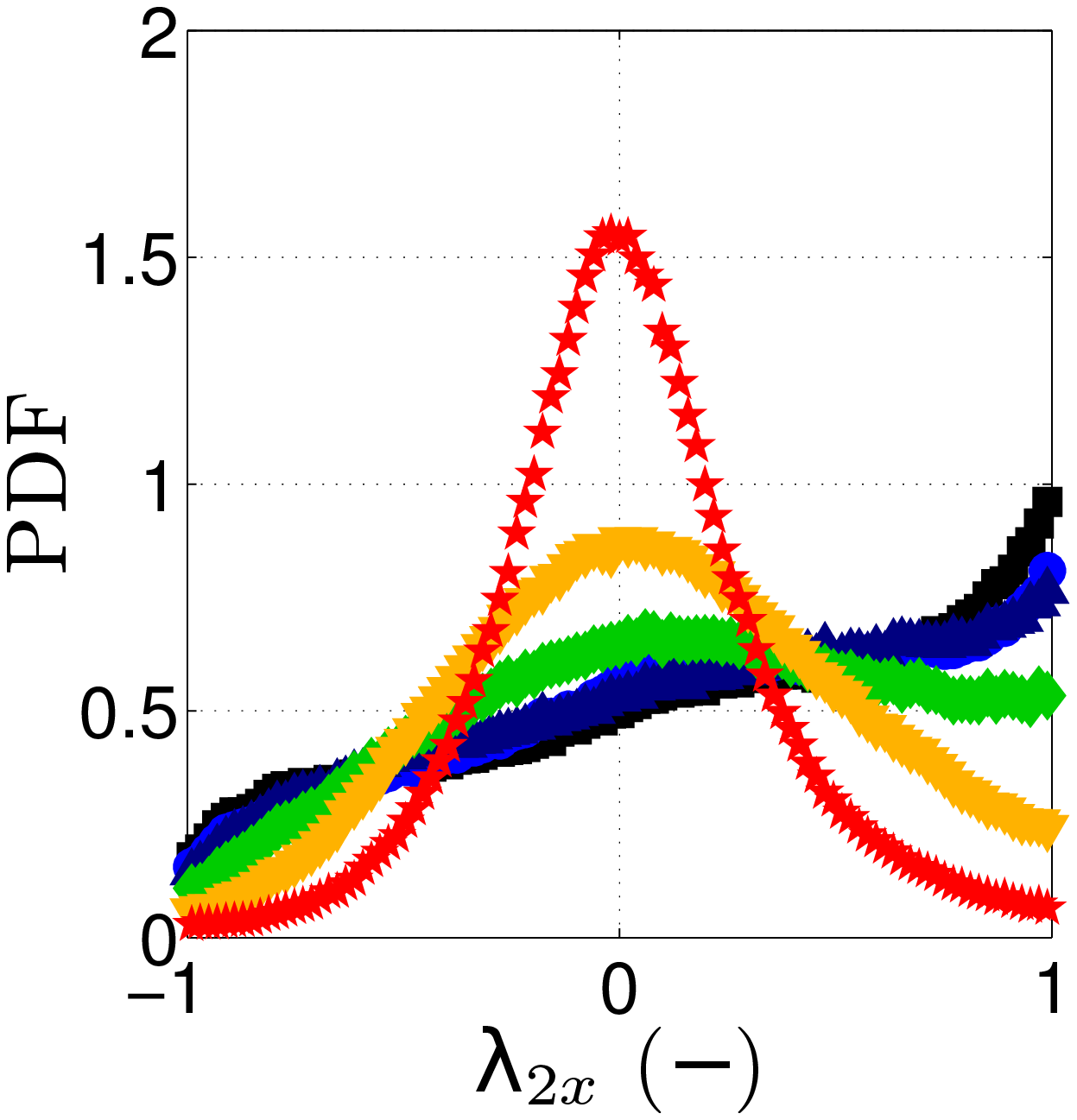}~%{exps123456_lambda2x-fstep30-linlin}
			\includegraphics[height = 0.205\textheight]{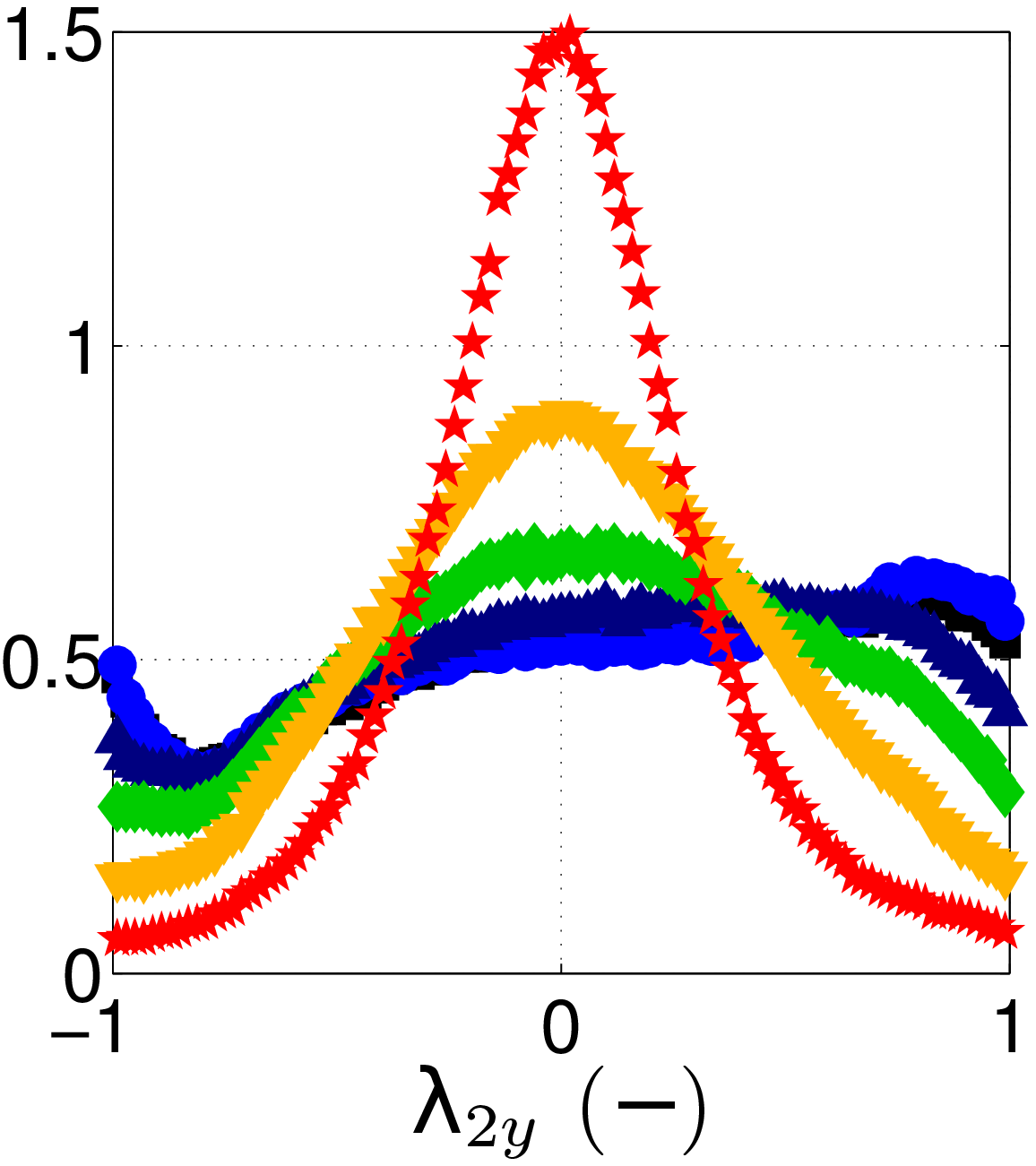}~%{exps123456_lambda2y-fstep30-linlin}
			\includegraphics[height = 0.205\textheight]{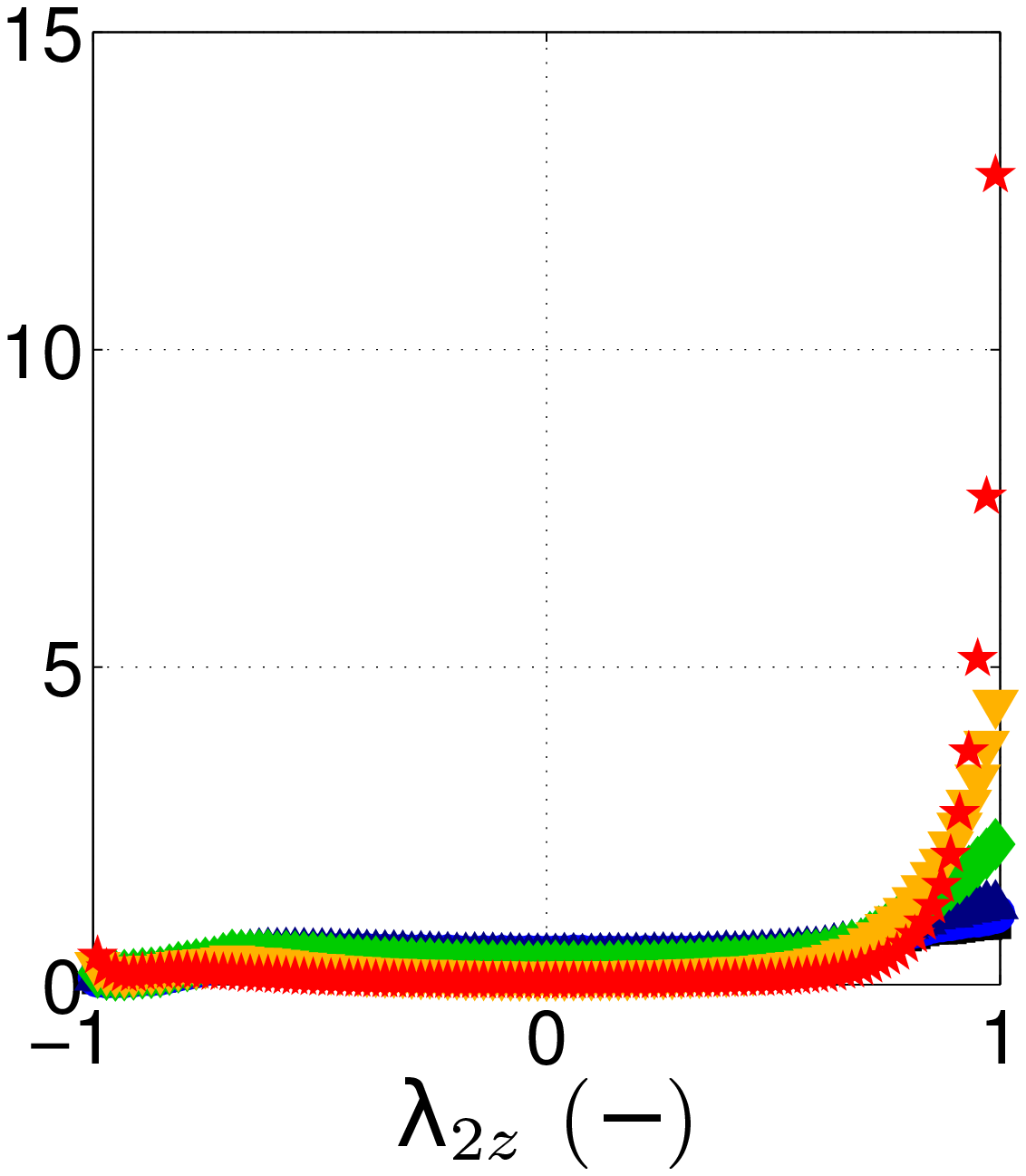}\\%{exps123456_lambda2z-fstep30-linlin}
			\includegraphics[height = 0.205\textheight]{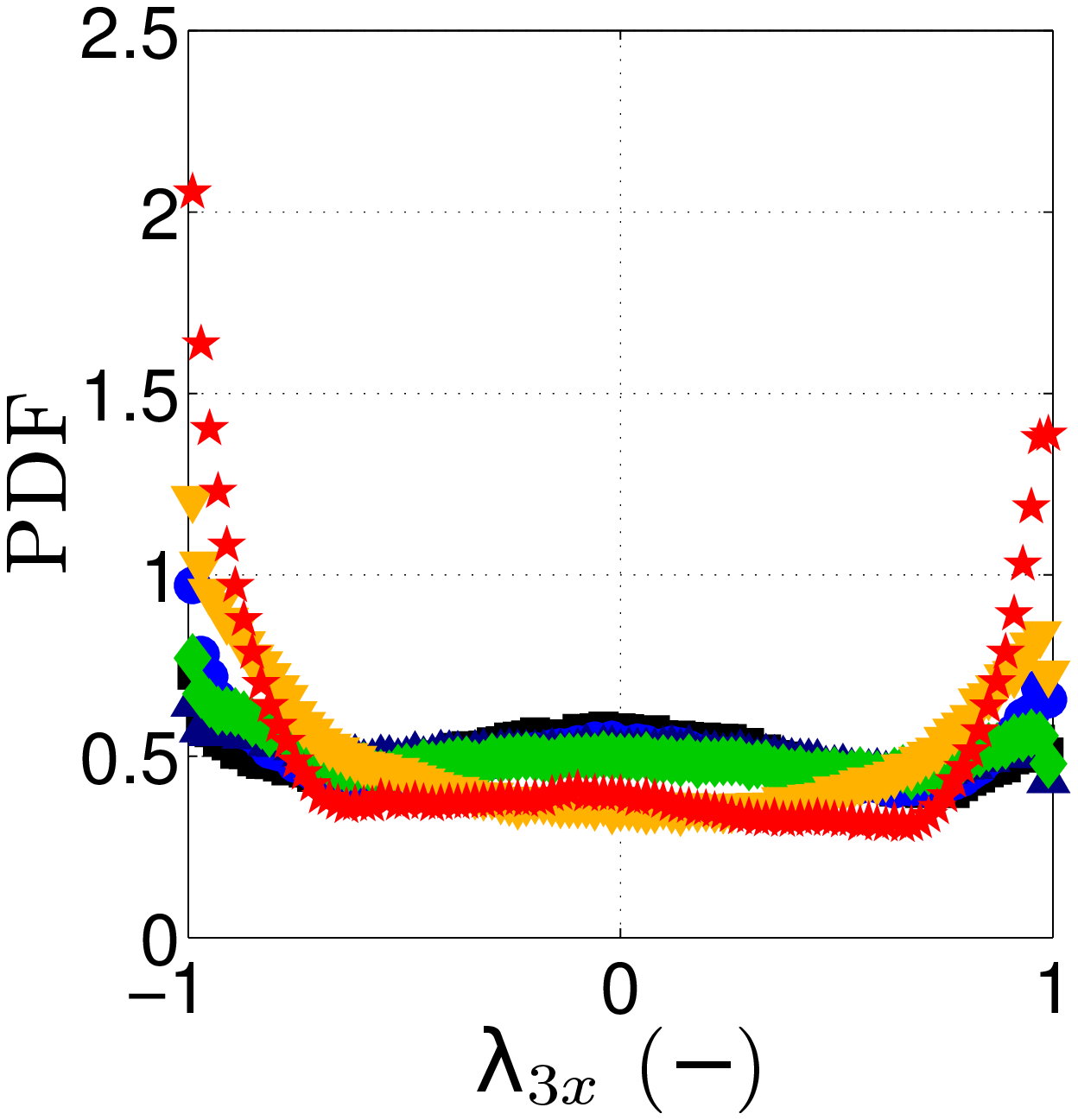}~%{exps123456_lambda3x-fstep30-linlin}
			\includegraphics[height = 0.205\textheight]{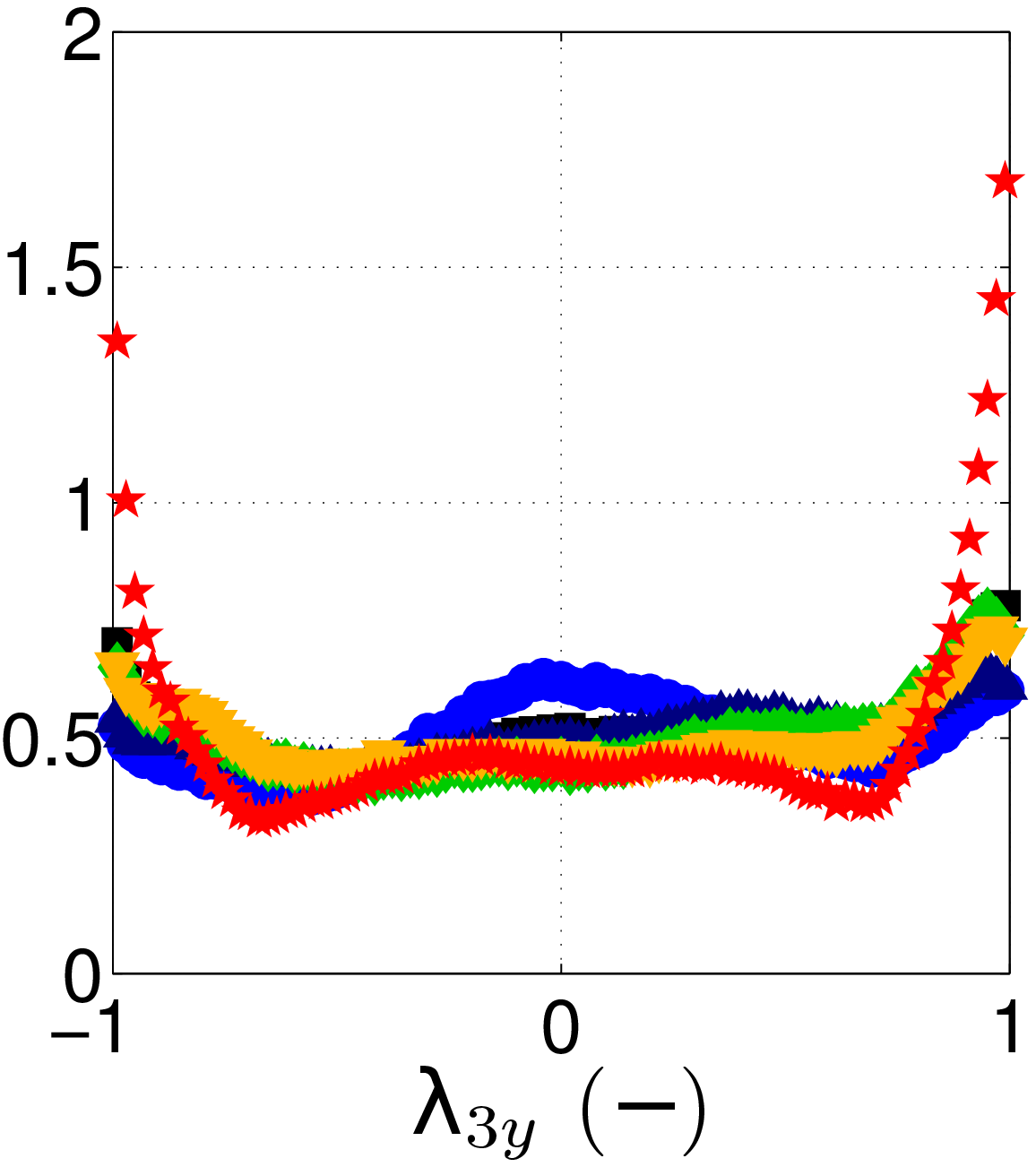}~%{exps123456_lambda3y-fstep30-linlin}
			\includegraphics[height = 0.205\textheight]{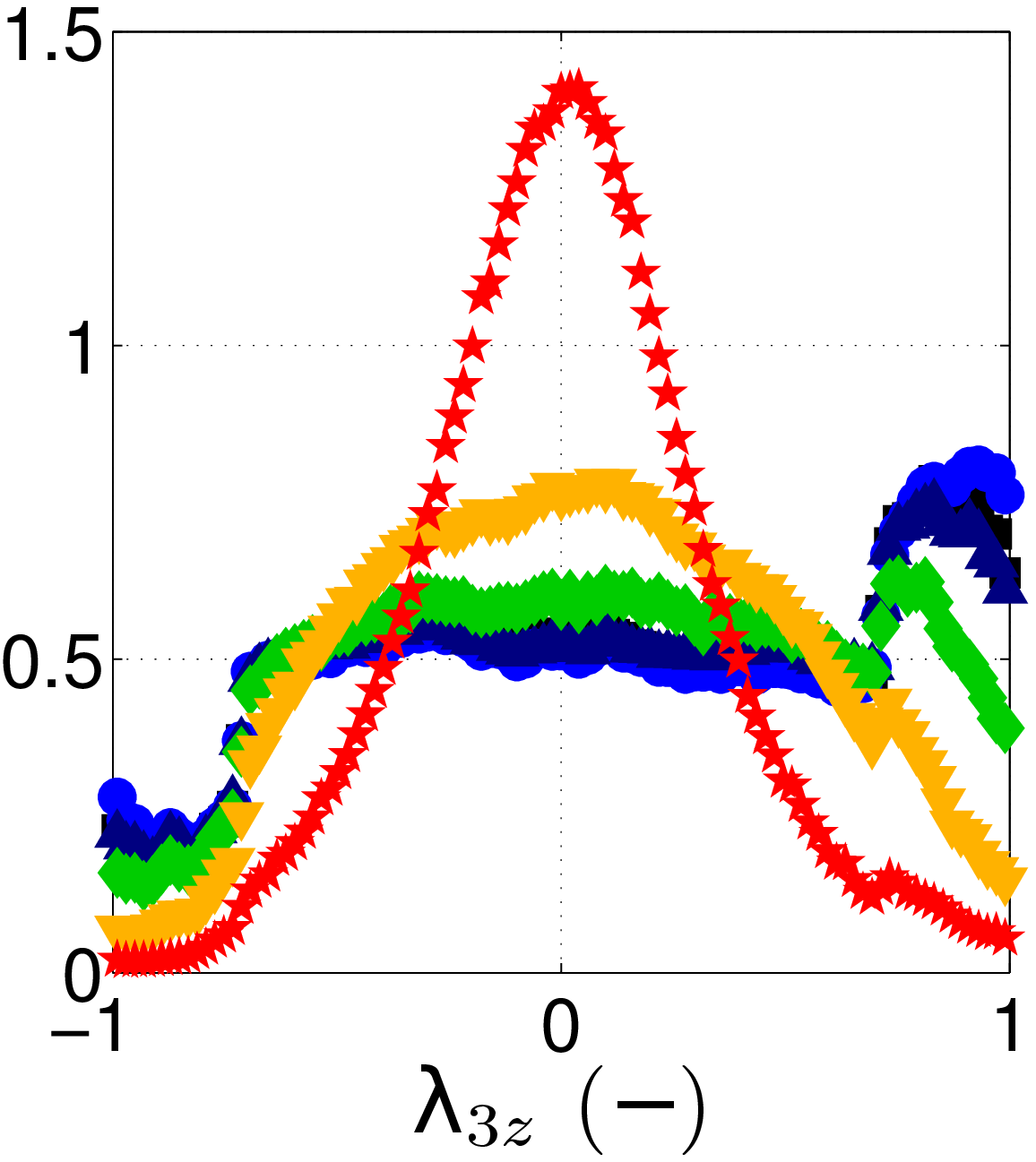}%{exps123456_lambda3z-fstep30-linlin}
			\caption{PDFs of the three Cartesian components $\lambda_{\alpha x}$, $\lambda_{\alpha y}$, $\lambda_{\alpha z}$ of each of the three eigenvectors ${\boldsymbol{\lambda}}_{\alpha}$ of the strain rate tensor, for all experiments (with $\Omega$ varying from 0 to 5 rad/s). Note the vertical axis of the PDF of $\lambda_{2z}$, which indicate much larger values for this PDF than observed in the remaining plots. Symbols as in Fig.~\ref{fig3}.}
			\label{fig6}
		\end{figure}
		In Fig.~\ref{fig6}, the Cartesian components of the eigenvectors ${\boldsymbol{\lambda}}_{\alpha}$ (with $\alpha=1, 2$ or 3) are presented, in order to investigate their alignment with respect to the rotation axis $z$.
		While the preferential alignment between ${\boldsymbol{\omega}}$ and ${\boldsymbol{\lambda}}_2$ gets progressively enhanced by increasing $\Omega$ (see Fig.~\ref{fig3}), the absolute value of the corresponding eigenvalue $\Lambda_2$ is strongly reduced by rotation (and symmetrically distributed around $\Lambda_2=0$), see the data for $\sigma_{\Lambda_2}$ and $S_{\Lambda_2}$ in Table \ref{tab2}. The strain rate eigenframe is seen to have almost no preferential orientation with respect to the Cartesian frame when no rotation is applied as the PDFs show more or less uniform distributions (a remark concerning $\lambda_{\alpha z}$ will follow below). With fast rotation, ${\boldsymbol{\lambda}}_2$ develops a strong preferential vertical alignment, while ${\boldsymbol{\lambda}}_1$ and ${\boldsymbol{\lambda}}_3$ get statistically horizontal.
		The more pronounced alignment between ${\boldsymbol{\omega}}$ and ${\boldsymbol{\lambda_2}}$ for high $\Omega$ is understood in terms of their mutual alignment with the vertical direction (parallel with ${\mathbf{\Omega}}$). The statistical orientation of $\lbrace{\boldsymbol{\lambda}}_1,{\boldsymbol{\lambda}}_2,{\boldsymbol{\lambda}}_3\rbrace$ with respect to $\lbrace x,y,z\rbrace$ is a signature of the transitional process from 3D- to Q2D-dynamics of the flow. The damping of the vertical gradients, together with the boundary conditions imposed by the fluid container, induce a strong reduction of the vertical velocity component (see Ref.~\cite{delcastello2011pre}).
		The flow evolves almost solely in the horizontal directions, and it is thus clear that also ${\boldsymbol{\lambda}}_1$ and ${\boldsymbol{\lambda}}_3$ (the stretching and compression axes of the strain rate tensor) lie -- in a statistical sense -- in the horizontal plane. For purely 2D-flows, $\Lambda_2$ vanishes and its eigenvector ${\boldsymbol{\lambda}}_2$ is perpendicular to the plane of motion, while the eigenvectors ${\boldsymbol{\lambda}}_1$ and ${\boldsymbol{\lambda}}_3$ lie in the plane.\\
		Finally, we would like to add two remarks. First, it is curious to note that, for zero background rotation, the flow is characterised by a slight tendency of horizontal alignment of the third eigenvector (${\boldsymbol{\lambda}}_3$). This indicates that compression in the flow happens more frequently in the $x$- or $y$-directions, rather than in the vertical one. This could be interpreted in view of the geometrical constraint of the flow in our experiment, which results from the aspect ratio of the fluid container $L_x:L_y:L_z=2:2:1$. Second, we do not fully understand the meaning of a preferential sign along a specific direction, which is observed in rotating and non-rotating runs for several PDFs shown above: see, \textit{e.g.}, the slight tendency of ${\boldsymbol{\lambda}}_1$ to align with the vertical upward (rather than downward) direction for no rotation, or the strong tendency of alignment between ${\boldsymbol{\lambda}}_2$ and the same vertical upward direction in case of fast rotation.
		Further investigation is being carried on in order to understand these issues.
	\section{Conclusions}\label{Sec7}
		We performed a series of experiments of confined and continuously forced turbulence subjected to background rotation. A range of rotation rates $\Omega$ was considered, from non-rotating to rapidly rotating turbulence with a maximum background rotation rate of $\Omega=5$ rad/s (with Rossby number much smaller than unity). Typically, the Taylor-scale Reynolds number in our experiments was around ${\rm{Re}}_{\lambda}\approx 100$. Based on the components of the strain rate tensor, measured in a Lagrangian way by 3D Particle Tracking Velocimetry, we have been able to quantify statistically the effects of system rotation on several flow properties. The experimental results show for the range of rotation rates considered how the turbulence evolves from almost isotropic 3D turbulence ($\Omega \lesssim 0.2$ rad/s) to quasi-2D turbulence ($\Omega\approx 5.0$ rad/s).
		The two--dimensionalisation process induced by rotation has been quantified in terms of the modified alignment -- in a statistical sense -- of the vorticity vector with respect to the eigenframe of the strain rate tensor and by analysing the projection of the vorticity vector ${\boldsymbol{\omega}}$ on the vortex stretching vector ${\bf{W}}$. When the rotation rate is increased, the probability of alignment of the vorticity vector with the intermediate eigenvector is amplified. At the same time, the same intermediate eigenvector tends to align statistically with the rotation axis, so that the first and third eigenvectors tend to lie in the plane of motion. This is consistent with the limit picture of 2D-turbulence, for which there exist only two eigenvectors (and they lie in the plane of motion) and the vorticity vector is perpendicular to the plane of motion. Therefore the vorticity vector is perpendicular to both eigenvectors, and no stretching of vorticity is possible.
		The latter phenomenon has been confirmed by exploring the PDFs of the geometrical invariant $\cos({\boldsymbol{\omega}},{\bf{W}})$ for the range of rotation rates considered in this study. The absence of the vortex stretching mechanism is one of the main distinguishing features between the dynamics of 3D- and 2D-turbulence. Finally, we have quantified the role of system rotation on the self-amplification terms of the enstrophy and strain rate equations and the direct contribution of the background rotation on these evolution equations. It turned out that the main effect of background rotation is a strong reduction of extreme events and a related strong reduction of the skewness of PDFs of several quantities, such as the intermediate eigenvalue of the strain rate tensor and the enstrophy and strain self-amplification terms.\\
		The present experimental investigations on rotating turbulence are carried out in a confined fluid with localised forcing (near the bottom part of the container). As a consequence, both the presence of Ekman boundary layers and of the forcing affect the dynamics of the turbulent flow, and the rms properties of the turbulent flow will show a decay in the vertical direction. This decay is substantial for the non- or weakly-rotating cases and relatively limited for rapidly rotating turbulence, see Ref.~\cite{bokhoven2009pof}. Statistically steady and homogeneous rotating turbulence is more easily achieved in DNS. Combination with state-of-the-art numerical studies on the Lagrangian statistics of (inertial) particles, previously mostly applied to homogeneous isotropic turbulence (see for a review Ref.~\cite{toschi2009}), may allow complementary studies on the role of rotation on geometrical statistics in rotating turbulence.
		It may also allow to cover a larger part of the $\{{\rm{Re}}_{\lambda},{\rm{Ro}}\}$ parameter space than in the present experiments.\\
		\\
		\textit{Acknowledgements:} This project has been funded by the Netherlands Organisation for Scientific Research (NWO) under the Innovational Research Incentives Scheme grant ESF.6239. The institutes IGP and IfU of ETH (Z\"{u}rich) are acknowledged for making available the PTV code. The European COST Action MP0806 ``Particles in Turbulence'' is also acknowledged.\\

\begin{thebibliography}{50}
		%\markboth{Taylor \& Francis and I.T. Consultant}{Journal of Turbulence}
		%
		\bibitem[1]{traugott1958naca}
		S.C. Traugott, {\em Influence of solid-body rotation on screen-produced turbulence}, Nat. Advis. Comm. Aero., Washington, Tech. Note 4135 (1958).
		\bibitem[2]{wigeland1978iit}
		R.A. Wigeland and H.M. Nagib, {\em Grid-generated turbulence with and without rotation about the streamwise direction, IIT Fluids and Heat Transfer}, R78-1, Illinois Inst. of Tech., Chicago, Illinois (1978).
		\bibitem[3]{jacquin1990jfm}
		L. Jacquin, O. Leuchter, C. Cambon, and J. Mathieu, {\em Homogeneous turbulence in the presence of rotation}, J. Fluid Mech. 220 (1990), pp. 1--52.
		\bibitem[4]{cambon1989jfm}
		C. Cambon and L. Jacquin, {\em Spectral approach to non-isotropic turbulence subjected to rotation}, J. Fluid Mech. 202 (1989), pp. 295--317.
		\bibitem[5]{ibbetson1975jfm}
		A. Ibbetson and D.J. Tritton, {\em Experiments on turbulence in a rotating fluid}, J. Fluid Mech. 68 (1975), pp. 639--672.
		\bibitem[6]{mcewan76}
		A.D. McEwan, {\em Angular momentum diffusion and the initiation of cyclones}, Nature 260 (1976), pp. 126--128. 
		\bibitem[7]{hopfinger1982jfm}
		E.J. Hopfinger, F.K. Browand, and Y. Gagne, {\em Turbulence and waves in a rotating tank}, J. Fluid Mech. 125 (1982), pp. 505--534.
		\bibitem[8]{yeung1998pof}
		P.K. Yeung and Y. Zhou, {\em Numerical study of rotating turbulence with external forcing}, Phys. Fluids 10 (1998), pp. 2895--2909.
		\bibitem[9]{godeferd1999jfm}
		F.S. Godeferd and L. Lollini, {\em Direct numerical simulations of turbulence with confinement and rotation}, J. Fluid Mech. 393 (1999), pp. 257--308.
		\bibitem[10]{baroud2003pof}
		C.N. Baroud, B.B. Plapp, H.L. Swinney, and Z.S. She, {\em Scaling in three-dimensional and quasi-two-dimensional rotating turbulent flows}, Phys. Fluids 15 (2003), pp. 2091--2104.
		\bibitem[11]{morize2005pof}
		C. Morize, F. Moisy, and M. Rabaud, {\em Decaying grid-generated turbulence in a rotating tank}, Phys. Fluids 17 (2005), 095105.
		\bibitem[12]{morize2006pof}
		C. Morize and F. Moisy, {\em Energy decay of rotating turbulence with confinement effects}, Phys. Fluids 18 (2006), 065107.
		\bibitem[13]{moisy2010jfm}
		F. Moisy, C. Morize, M. Rabaud, and J. Sommeria, {\em Decay laws, anisotropy and cyclone-anticyclone asymmetry in decaying rotating turbulence}, J. Fluid Mech. 666 (2011), pp. 5--35.
		\bibitem[14]{davidson2006jfm}
		P.A. Davidson, P.J. Staplehurst, and S.B. Dalziel, {\em On the evolution of eddies in a rapidly rotating system}, J. Fluid Mech. 557 (2006), pp. 135--144.
		\bibitem[15]{bokhoven2009pof} 
		L.J.A. {van {Bokhoven}}, H.J.H. Clercx, G.J.F. {van {Heijst}}, and R.R. Trieling, {\em Experiments on rapidly rotating turbulent flows}, Phys. Fluids 21 (2009), 096601.
		\bibitem[16]{thalabard2011}
		S. Thalabard, D. Rosenberg, A. Pouquet, and P.D. Mininni, {\em Conformal invariance in three-dimensional rotating turbulence}, Phys. Rev. Lett. 106 (2011), 204503.
		\bibitem[17]{mininni2012} 
		P.D. Mininnia, D. Rosenberg, and A. Pouquet, {\em Isotropization at small scales of rotating helically driven turbulence}, J. Fluid Mech. 699 (2012), pp. 263--279.
		\bibitem[18]{kinzel2010}
		M. Kinzel, M. Wolf, M. Holzner, B. L{\"{u}}thi, C. Tropea, and W. Kinzelbach, {\em Simultaneous two-scale 3D-PTV measurements in turbulence under the influence of system rotation}, Exp. Fluids 51 (2010), pp. 75--82.
		\bibitem[19]{delcastello2011pre}
		L. Del Castello and H.J.H. Clercx, {\em Lagrangian velocity autocorrelations in statistically steady rotating turbulence}, Phys. Rev. E 83 (2011), 056316.
		\bibitem[20]{delcastello2011prl}
		L. Del Castello and H.J.H. Clercx, {\em Lagrangian acceleration of passive tracers in statistically--steady rotating turbulence}, Phys. Rev. Lett. 107 (2011), 214502.
		\bibitem[21]{luethi2005jfm}
		B. L{\"u}thi, A. Tsinober, and W. Kinzelbach, {\em Lagrangian measurement of vorticity dynamics in turbulent flow}, J. Fluid Mech. 528 (2005), pp. 87--118.
		\bibitem[22]{luethi2008ercoftac}
		B. L{\"u}thi, M. Kinzel, A. Liberzon, M. Holzner, C. Tropea, W. Kinzelbach, and A. Tsinober, {\em 3d-2d transition in inhomogeneous rotating turbulent flow}, ERCOFTAC Special Bulletin on Environmental Flows (2008).
		\bibitem[23]{mininni2009pof}
		P.D. Mininni, A. Alexakis, and A. Pouquet, {\em Scale interactions and scaling laws in rotating flows at moderate Rossby numbers and large Reynolds numbers}, Phys. Fluids 21 (2009), 015108.
		\bibitem[24]{delcastello2010phd}
		L. Del Castello,  {\em Table--top rotating turbulence: an experimental insight through Particle Tracking}, Ph.D. diss., Eindhoven University of Technology, The Netherlands, 2010.
		\bibitem[25]{taylor1938}
		G.I. Taylor, {\em Production and dissipation of vorticity in a turbulent fluid}, Proc. Roy. Soc. London A164 (1938), pp. 15--23.
		\bibitem[26]{tsinober2001}
		A. Tsinober, {\itshape An informal introduction to turbulence}, Kluwer Academic, Dordrecht, 2001.
		\bibitem[27]{galantitsinober2000}
		B. Galanti and A. Tsinober, {\em Self-amplification of the field of velocity derivatives in quasi-isotropic turbulence}, Phys. Fluids 12 (2000), pp. 3097--3099.
		\bibitem[28]{saffmann1991}
		P.G. Saffman, in {\it The Global Geometry of Turbulence}, J. Jimenez, Plenum, eds., NATO ASI Ser. B 268, 1991, p. 348.
		\bibitem[29]{tsinober1992}
		A. Tsinober, E. Kit, and T. Dracos, {\em Experimental investigation of the field of velocity gradients in turbulent flows}, J. Fluid Mech. 242 (1992), pp. 169--192.
		\bibitem[30]{tsinober1996}
		A. Tsinober, {\em Geometrical statistics in turbulence}, Advances in Turbulence 6 (1996), pp. 263--266.
		\bibitem[31]{kholmyansky2001pof}
		M. Kholmyansky, A. Tsinober, and S. Yorish, {\em Velocity derivatives in the atmospheric turbulent flow at $Re_{\lambda}=10^4$},  Phys. Fluids 13 (2001), pp. 311--314.
		\bibitem[32]{guala2005jfm}
		M. Guala, B. L{\"u}thi, A. Liberzon, A. Tsinober, and W. Kinzelbach, {\em On the evolution of material lines and vorticity in homogeneous turbulence}, J. Fluid Mech. 533 (2005), pp. 339--359.
		\bibitem[33]{sommeria1986}
		J. Sommeria, {\em Experimental study of the two-dimensional inverse energy cascade in a square box}, J. Fluid Mech. 170 (1986), pp. 139--68.
		\bibitem[34]{tabeling1991}
		P. Tabeling, S. Burkhart, O. Cardoso, and H. Willaime, {\em Experimental study of freely decaying two-dimensional turbulence}, Phys. Rev. Lett. 67 (1991), n. 27, pp. 3772--3775.
		\bibitem[35]{dolzhanskii1992}
		F.V. Dolzhanskii, V.A. Krymov, and D.Y. Manin, {\em An advanced experimental investigation of quasi-two-dimensional shear flows}, J. Fluid Mech. 241 (1992), pp. 705--722.
		\bibitem[36]{clercx2003}
		H.J.H. Clercx, G.J.F. {van {Heijst}}, and M.L. Zoeteweij, {\em Quasi-two-dimensional turbulence in shallow fluid layers: The role of bottom friction and fluid layer depth}, Phys. Rev. E 67 (2003), 066303.
		\bibitem[37]{akkermans2008}
		R.A.D. Akkermans, A.R. Cieslik, L.P.J. Kamp, R.R. Trieling, H.J.H. Clercx, and G.J.F. {van Heijst}, {\em The three-dimensional structure of an electromagnetically generated dipolar vortex in a shallow fluid layer}, Phys. Fluids 20 (2008), 116601.
		\bibitem[38]{maas1993eif1}
		H.G. Maas, {\em High-speed solid state camera systems for digital photogrammetry}, Int. Arch. Photogrammetry and Remote Sensing 29 (1993), pp. 482--482.
		\bibitem[39]{malik1993eif2}
		N.A. Malik, T. Dracos, and D.A. Papantoniou, {\em Particle tracking velocimetry in three-dimensional flows. Part II: Particle tracking}, Exp. Fluids 15 (1993), pp. 279--294.
		\bibitem[40]{willneff2002istpdrm}
		J. Willneff and A. Gruen, {\em A new spatio-temporal matching algorithm for 3D-particle tracking velocimetry}, Proc. 9th Int. Symp. Transport Phenomena and Dynamics of Rotating Machinery 10 (2002), p. 14.
		\bibitem[41]{willneff2003phd}
		J. Willneff,  {\em A spatio-temporal matching algorithm for 3D particle tracking velocimetry}, Ph.D. diss., Swiss Federal Institute of Technology Z{\"u}rich, 2003.
		\bibitem[42]{luthi2002phd}
		B. L{\"u}thi,  {\em Some aspects of strain, vorticity and material element dynamics as measured with 3D particle tracking velocimetry in a turbulent flow}, Ph.D. diss., Swiss Federal Institute of Technology Z{\"u}rich, 2002.
		\bibitem[43]{kloosterziel1991}
		R.C. Kloosterziel and G.J.F. van Heijst, {\em An experimental study of unstable barotropic vortices in a rotating fluid}, J. Fluid Mech. 223 (1991), pp. 1--24.
		\bibitem[44]{vanheijst2009}
		G.J.F. van Heijst and H.J.H. Clercx, {\em Laboratory modeling of geophysical vortices}, Annu. Rev. Fluid Mech. 41 (2009), pp. 143--164.
		\bibitem[45]{siggia1981}
		E.D. Siggia, {\em Numerical study of small-scale intermittency in three-dimensional turbulence}, J. Fluid Mech. 107 (1981), pp. 375--406.
		\bibitem[46]{ashurst1987}
		W.T. Ashurst, A.R Kerstein, R.A. Kerr, and C.H. Gibson, {\em Alignment of vorticity and scalar gradient with strain rate in simulated Navier-Stokes turbulence}, Phys. Fluids 30 (1987), pp. 2343--2353.
		\bibitem[47]{tsinober1998a}
		A. Tsinober, {\em Is concentrated vorticity that important?}, Eur. J. Mech. B/Fluids 17 (1998), pp. 421--449.
		\bibitem[48]{betchov1956jfm}
		R. Betchov, {\em An inequality concerning the production of vorticity in isotropic turbulence}, J. Fluid Mech. 1 (1956), pp. 497--504.
		\bibitem[49]{luethi2007jot}
		B. L{\"u}thi, S. Ott, J. Berg, and J. Mann, {\em Lagrangian multi-particle statistics}, J. Turbulence 8 (2007), 45.
		\bibitem[50]{toschi2009}
		F. Toschi and E. Bodenschatz, {\em Lagrangian properties of particles in turbulence}, Annu. Rev. Fluid Mech. 41 (2009), pp. 375--404. 
		%
		%\markboth{Taylor \& Francis and I.T. Consultant}{Journal of Turbulence}
	\end{thebibliography}
\end{document}